\documentclass[10pt,journal,compsoc]{IEEEtran}

\makeatletter
\AddToHook{cmd/added/before}{\color{black}}
\makeatother
\usepackage{amsmath,amsfonts,amssymb} 
\usepackage{algorithmic}
\usepackage{array}
\usepackage{textcomp}
\usepackage{stfloats}
\usepackage{url} 
\usepackage{verbatim}
\usepackage{graphicx}
\def\BibTeX{{\rm B\kern-.05em{\sc i\kern-.025em b}\kern-.08em
    T\kern-.1667em\lower.7ex\hbox{E}\kern-.125emX}}
\usepackage{balance} 
\usepackage[table,xcdraw]{xcolor} 
\usepackage{colortbl} 
\usepackage{tcolorbox}
\usepackage{listings}
\usepackage{makecell} 

\definecolor{ashgrey}{rgb}{0.98, 0.91, 0.71}
\definecolor{grey}{rgb}{0.9, 0.9, 0.9}

\usepackage{multirow} 
\usepackage{hyperref}
\usepackage{booktabs}
\usepackage[utf8]{inputenc}
\usepackage{calligra}
\usepackage[normalem]{ulem}
\usepackage{indentfirst}
\usepackage{enumitem}
\usepackage{nicematrix}
\usepackage{fontawesome5}
\usepackage[ruled,vlined]{algorithm2e}

\usepackage[commandnameprefix=ifneeded]{changes} 

\usepackage{threeparttable}
\usepackage{subcaption}
\usepackage{float}
\usepackage{amsthm}
\theoremstyle{definition}
\newtheorem{definition}{Definition}[section]

\definecolor{codegreen}{rgb}{0,0.6,0}
\definecolor{codegray}{rgb}{0.5,0.5,0.5}
\definecolor{codepurple}{rgb}{0.58,0,0.82}
\definecolor{backcolour}{rgb}{0.95,0.95,0.92}
\definecolor{cerulean}{rgb}{0.0, 0.48, 0.65}
\definecolor{ceruleanblue}{rgb}{0.16, 0.32, 0.75}
\definecolor{cadmiumred}{rgb}{0.89, 0.0, 0.13}

\definecolor{viol}{RGB}{134,0,175}
\definecolor{lightgreen}{RGB}{217,242,208}
\definecolor{lightred}{RGB}{252,239,241}
\definecolor{darkgreen}{RGB}{113,216,148}
\definecolor{darkred}{RGB}{229, 168, 172}

\DeclareMathAlphabet{\mathcalligra}{T1}{calligra}{m}{n}

\newcommand{\ocddata}{\textsc{OcdData}\xspace}
\newcommand{\jitdata}{\textsc{JITData}\xspace}

\newcommand{\name}{\textsc{CCISolver}\xspace}
\newcommand{\detector}{\textsc{CCIDetector}\xspace}
\newcommand{\fixer}{\textsc{CCIFixer}\xspace}
\newcommand{\bench}{\textsc{CCIBench}\xspace}

\DeclareUnicodeCharacter{2212}{-}
\begin{document}
\title{\name: End-to-End Detection and Repair of Method-Level Code-Comment Inconsistency}

\author{
    Renyi Zhong, Yintong Huo, Wenwei Gu, Jinxi Kuang, Zhihan Jiang, Guangba Yu, Yichen Li, \\ David Lo , and
    Michael R. Lyu \\
    \IEEEcompsocitemizethanks{\IEEEcompsocthanksitem Renyi Zhong, Wenwei Gu, Jinxi Kuang, Zhihan Jiang, Guangba Yu, Yichen Li, Michael R. Lyu are with The Chinese University of Hong Kong, Hong Kong. (e-mail:ryzhong22@cse.cuhk.edu.hk;~wwgu21@cse.cuhk.edu.hk;~jxkuang22\\@cse.cuhk.edu.hk;~zhjiang22@cse.cuhk.edu.hk;~guangbayu@cuhk.edu.hk;~ycli21@cse.cuhk.edu.hk;~lyu@cse.cuhk.edu.hk)
    \IEEEcompsocthanksitem Yintong Huo and David Lo are with Singapore Management University, Singapore. (e-mail: ythuo@smu.edu.sg; davidlo@smu.edu.sg)}
}

\markboth{Manuscript}{XXXX}

\maketitle

\begin{abstract}

\added{
Comments within code serve as a crucial foundation for software documentation, facilitating developers to communicate and understand the code effectively. However, code-comment inconsistency (CCI) can negatively affect software development, testing, and maintenance. Recent efforts to mitigate this issue have emerged, but existing studies often suffer from inaccurate datasets and inadequate solutions, weakening their practical effectiveness.
In this study, we first conduct a quantitative analysis of existing datasets, revealing a substantial portion of sampled data are mislabeled. To address these data limitations, we introduce \bench, a refined dataset comprising high-quality data, to support the training and evaluation of method-level CCI methods. Furthermore, we present an innovative end-to-end LLM-based framework, \name, designed to improve code quality by identifying and rectifying CCIs. 
Comprehensive evaluations demonstrate \name's superior performance. For detection, it establishes a new state-of-the-art with an F1-score of 89.54\%. In fixing task, it achieves a remarkable 18.84\% relative improvement in GLEU score over the strongest baseline. This superiority is confirmed by human evaluation, where \name's fixing success rate of 0.6533 significantly surpasses existing methods. Critically, in a practical end-to-end setting, \name's innovative architecture is approximately 36\% faster for inference than the baseline model, underscoring its scalability and real-world applicability.
}

\end{abstract}


\vspace{0.4in}

\vspace{-35pt}
\section{Introduction}


\IEEEPARstart{C}{ode} comments serve as fundamental documentation artifacts in software development, providing crucial metadata and explanations that facilitate program comprehension and maintenance. These textual annotations enable developers to communicate essential information, including implementation details, component relationships, and code evolution rationale~\cite{pascarella2019classifying,padioleau2009listening}. Research has demonstrated that well-maintained comments can enhance software maintainability and support developers' understanding of complex codebases~\cite{buse2009learning}.

\begin{figure} [t]
    \centering
    \includegraphics[width=0.9\linewidth]{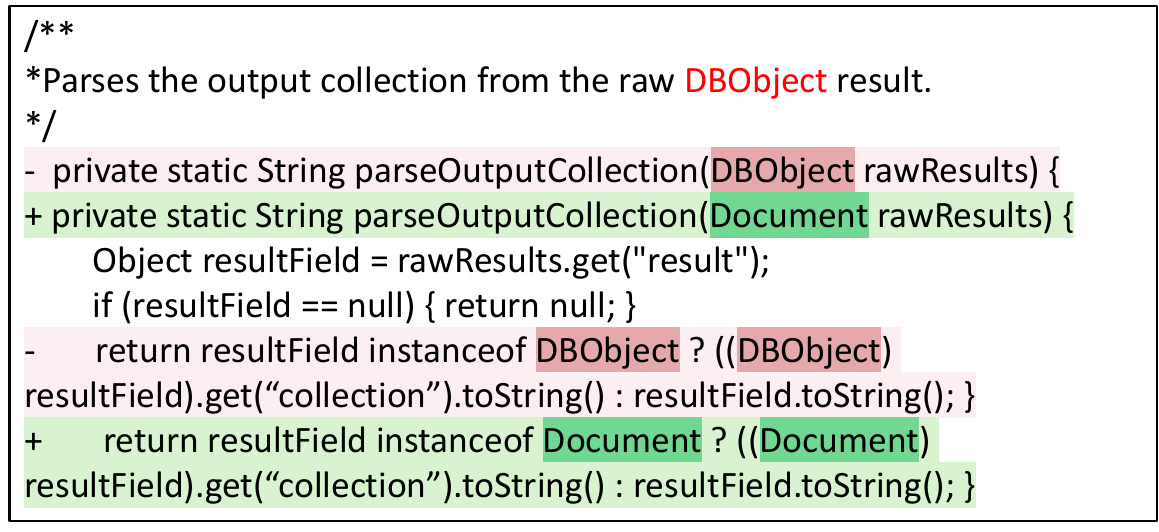}
    \caption{A CCI example in spring-data-mongodb~\cite{springdatamongodb}.}
    \label{fig:intro_example}
    \vspace{-0.2in}
\end{figure}

To ensure that comments align with their related code, it is imperative for comments to be updated alongside code modifications~\cite{zhong2025toward}. However, developers often neglect to revise comments during code changes or refactoring, resulting in code-comment inconsistencies (CCI) where comments misrepresent the code. For instance, Figure~\ref{fig:intro_example} illustrates a CCI scenario in spring-data-mongodb, where a method was modified to handle Document objects, yet the comment incorrectly claims it works with DBObject. These inconsistencies can hinder development workflows by misrepresenting code intent and wasting developers' time interpreting incorrect comments~\cite{liu2023JustInTime}.

Recognizing the scale of modern systems and the labor-intensive nature of manually resolving CCI, researchers have proposed automated CCI detection techniques. These techniques leverage both traditional machine learning classifiers~\cite{liu2018Automatic} and advanced deep learning models~\cite{liu2023JustInTime,panthaplackel2021Deepa,xu2023Data}. However, there are limitations from both dataset side and proposed solution side. Specifically, existing datasets contain data mislabelling issues originating from the trade-offs made during dataset construction~\cite{xu2023Data}. Existing deep learning methods~\cite{liu2023JustInTime,panthaplackel2021Deepa} excel in detecting CCIs but are inherently limited in fixing them, while approaches like C4RLLaMA~\cite{rong2024code} that utilize fine-tuned LLMs for both tasks introduce significant computational overhead, particularly during the detection phase.


\textbf{Existing dataset limitations.} The current CCI datasets (e.g., \jitdata~\cite{panthaplackel2021Deepa} and \ocddata~\cite{liu2023JustInTime}) contain data mislabeling issues stemming from the trade-offs made during their construction. Researchers often opt for automated labeling over manual labeling due to the latter's high costs, although it can lead to labeling errors~\cite{xu2023Data,xu2024Code}. 
Specifically, inconsistencies are identified by comparing pairs of methods and comments from older and newer code versions, under the assumption that an individual comment update indicates a correction for the method, thus labeling the old comment as inconsistent with the new code. However, this process relies on an overly broad assumption that may not hold true in practice. 
For example, a comment may be updated to correct typos or add information, which could mistakenly classify it as inconsistent. While previous studies~\cite{rong2024code,xu2024Code} have acknowledged the issue of erroneous labeling, no further investigation has been conducted to identify the causes of false positives and their prevalence. 

\textbf{Existing solution limitations.} A complete CCI solution requires both detecting inconsistencies and fixing them by suggesting appropriate comment revisions~\cite{rong2024code}. While existing deep learning methods~\cite{liu2023JustInTime,panthaplackel2021Deepa} have shown promising result in detection phase, they fall short in fixing CCIs.
Recent advancements in large language model (LLM) have demonstrated exceptional capabilities in code understanding and repair~\cite{lyu2024automatic}. C4RLLaMA~\cite{rong2024code} utilizes a fine-tuned LLM to handle both detection and correction activities. Nonetheless, this method incurs considerable performance overhead because the detection task examines a vast number of cases, most of which do not require further inconsistency fixing.

\added{To investigate the potential dataset limitations, we first conduct a preliminary study on the widely-used \jitdata dataset. Our findings are alarming: a manual inspection of 600 sampled cases revealed that a staggering 45.67\% of instances labeled as positive (i.e., inconsistent) are, in fact, incorrectly identified (§~\ref{sec:study}). We pinpoint four primary reasons for these errors: \textit{add/delete information}, \textit{fix typos}, \textit{change case}, and \textit{change lexical}. These results not only alarm critical data quality issues in existing resources, but also underscore the urgent need for a cleaner dataset.}

\added{In response to the data quality issues identified, we develop a new dataset, \bench, derived from \jitdata. The construction of \bench aims to directly address the identified shortcomings, such as high false positive rates and their underlying causes, to provide a high-quality resource for CCI detection and fixing tasks. This involved a multi-stage process incorporating advanced de-duplication, syntactic cleaning, and semantic-based filtering. Initially, we eliminate duplicates in the data, a common issue that inflates existing datasets~\cite{xu2023Data}. We then use refined syntactic rules to remove trivial changes (e.g., typo corrections, case adjustments) that do not lead to CCI, directly addressing some of the previously identified false positive sources. To address more complicated cases, particularly those related to changes like information addition or deletion which syntactic rules may miss, we employ three powerful LLMs (i.e., GPT4o~\cite{OpenAIGPT4o}, Claude3.5-Sonnet~\cite{anthropic2024claude}, and LLaMA3.1-405b~\cite{llama3modelcard}) as independent voters to identify true CCIs, resulting in a dataset of 22,360 verified CCI cases.}

\added{Beyond the critical need for a cleaner dataset, effectively addressing CCIs also requires overcoming the limitations inherent in current solutions. As previously discussed, existing approaches either excel at detection but lack repair capabilities, or utilize computationally intensive LLMs for both tasks, leading to inefficiencies. To bridge this gap, we propose \name, an innovative end-to-end framework designed for both efficient CCI detection and effective automated repair. \name combines the strengths of different methodologies: (1) For detection, the component \detector uses a traditional deep learning approach. It first extracts code comments and associated code changes, then encodes this information for a semantic-based classifier to accurately identify inconsistent cases. This ensures efficient processing of numerous instances, a crucial aspect where LLM-only detection can be prohibitively slow and costly. (2) Once an inconsistency is detected, the \fixer component leverages the generative power of LLMs for repair. Specifically, \fixer is fine-tuned on the high-quality inconsistent cases derived from our newly developed \bench dataset. This targeted fine-tuning enables it to learn robust correction patterns and generate accurate and contextually appropriate comment updates.}

We conduct comprehensive evaluations of CCI detection methods using both \jitdata and our newly developed \bench dataset. Our evaluation reveals the following key findings: (1) Our \detector component achieves state-of-the-art performance in CCI detection. With an F1 score of 89.54\%, it outperforms existing methods by 1.73\%, establishing itself as the leading approach for CCI detection. (2) Our \fixer component demonstrates remarkable effectiveness in comment repair. It substantially outperforms existing solutions, achieving GLEU score improvements of 18.84\% and 12.07\% on the full and validated test sets, respectively. Human evaluation further validates these results, with \fixer achieving a success rate of 0.6533, significantly surpassing both DeepJIT (0.4867) and C4RLLaMA (0.5867). \added{The end-to-end evaluation confirms that the two-stage architecture of \name is both more effective and time efficient, improving detection and fixing quality while reducing inference time by approximately 36\%, making it a more practical and scalable solution.} The above results demonstrate the supreme quality of \bench and the effectiveness of \name in handling CCI issues.

\textbf{Contributions.} We conclude this paper's contribusions:
\begin{itemize}[leftmargin=*]
    \item \added{\textbf{[Study]} First quantitative analysis of label quality about CCI. Our study represents the first quantitative investigation into CCI data, uncovering potential issues with label quality. We identify four primary categories of labeling errors: addition/deletion of information, typo corrections, changes in case, and alterations in lexical content.}
    \item \added{\textbf{[Dataset]} Construction of a high-quality CCI dataset, \bench. We build \bench, a new dataset for CCI research. Its construction involves a multi-stage pipeline that combines syntactic cleaning to filter trivial changes and a novel semantic filtering process.}
    \item \added{\textbf{[Framework]} \name: A novel and efficient two-stage framework. We propose \name, a novel framework for efficient CCI detection and repair. It features a two-stage architecture: a lightweight deep learning-based \detector for fast and accurate detection , and a powerful LLM-based \fixer to repair inconsistencies. This decoupled design is more practical and significantly faster than monolithic LLM baselines in an end-to-end setting.}
    \item \added{\textbf{[Evaluation]} Comprehensive empirical validation. Extensive experiments show that \name surpasses existing methods in both CCI detection and repair. Its effectiveness is validated by automated metrics and rigorous human evaluation, which confirms its superior ability to generate correct fixes. The design also provides an advantage in time efficiency, making it a more practical and scalable solution.}
\end{itemize}

\vspace{-10pt}
\section{Background and Motivation}
\label{preliminarystudy}
\label{motivation}
\subsection{Code-Comment Inconsistency}

\subsubsection{Definitions}
We first clarify some frequently used concepts in this work. 

\begin{definition}[Code-comment pair] A code-comment pair ($M$, $C$) is comprised of a method $M$, which encapsulates a functional segment, and an associated method comment $C$ that pertains specifically to the method $M$. Note that we only consider method comments, excluding class comments and inline comments.
\end{definition}

\begin{definition}[Code Comment Inconsistency] A code comment inconsistency between a code-comment pair $(M, C)$ is described as 
\begin{equation}
\text{Inconsistency}(M, C) =
\begin{cases}
1, & \text{if } f(M{\prime}) \neq g(C), \\
& \text{where } M \to M{\prime}; \\
0, & \text{otherwise.}
\end{cases}
\label{eq:inconsistency}
\end{equation}
where $f(M{\prime})$ denotes the factual behavior or operation of the modified code $M{\prime}$, and $g(C)$ signifies the intention or interpretation of the original comment $C$. The condition $f(M{\prime}) \neq g(C)$ indicates that the revised code’s behavior is inconsistent with the comment’s meaning. The notation $M \to M{\prime}$ indicates the code change.
\label{de:inconsistency}
\end{definition}

\begin{definition}[CCI Detection Task]
Given a code change ($M \to M{\prime}$) and a related comment $C$, the CCI detection task is to determine whether the $M \to M{\prime}$ lead to a CCI according to the definition in equation~\ref{eq:inconsistency}. The output of this task is a binary classification indicating whether the inconsistency exist or not.

\end{definition}

\begin{definition}[CCI Fix Task]
Given a code change ($M \to M{\prime}$) and a related comment $C$, where $M{\prime}$ is inconsistent with $C$, the CCI fix task aims to generate a modified comment $C{\prime}$ that ensures $f(M{\prime}) = g(C{\prime})$.

\end{definition}



\subsubsection{Two scenarios}
Following prior work~\cite{panthaplackel2021Deepa}, we categorize this CCI task into two distinct scenarios.

\textit{Just-in-time (JIT)}: The objective here is to detect inconsistencies at the moment code changes occur, prior to their commitment. Specifically, $M$, $M{\prime}$, and $C$ are accessible, enabling an examination of the differences between the old and new versions.

\textit{Post-hoc}: In this case, only a single version of the code comment pair ($M$, $C$) is available for assessment. 

In this study, we concentrate on the just-in-time (JIT) scenario, as it closely aligns with the practical needs of contemporary software development. Identifying mismatches at the point of code modifications enables immediate resolution of potential issues before they are committed, thereby minimizing the likelihood of introducing errors into the codebase. Furthermore, the JIT setting is also adopted by
recent studies on CCI detection~\cite{xu2023Data,rong2024code}, highlighting its pivotal role in current development processes.

\subsubsection{Existing dataset}
\jitdata~\cite{panthaplackel2021Deepa} is created by identifying inconsistencies in code changes in the commit histories of 1,518 open-source Java projects. The dataset encompasses positive instances where comments are updated along with code changes, and negative instances where comments remain unchanged despite code modifications. Heuristics~\cite{panthaplackel2020associating} are employed to filter out irrelevant comment updates, while concentrating on @return, @param, and summary comments. A curated sample of 300 cases is chosen for reliable evaluation, with negative instances downsampled to ensure a balanced dataset. Table~\ref{tab:dataset_statistics} presents the core statistics of \jitdata.

\begin{figure}[htbp]
    \centering
    \subfloat[The example of return inconsistency.]{%
        \includegraphics[width=0.8\linewidth]{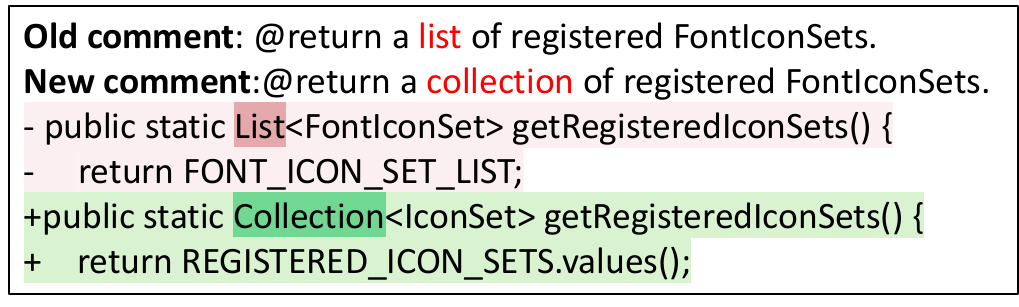}
        \label{fig:example1}
    }
    \hfill
    \subfloat[The example of param inconsistency.]{%
        \includegraphics[width=0.8\linewidth]{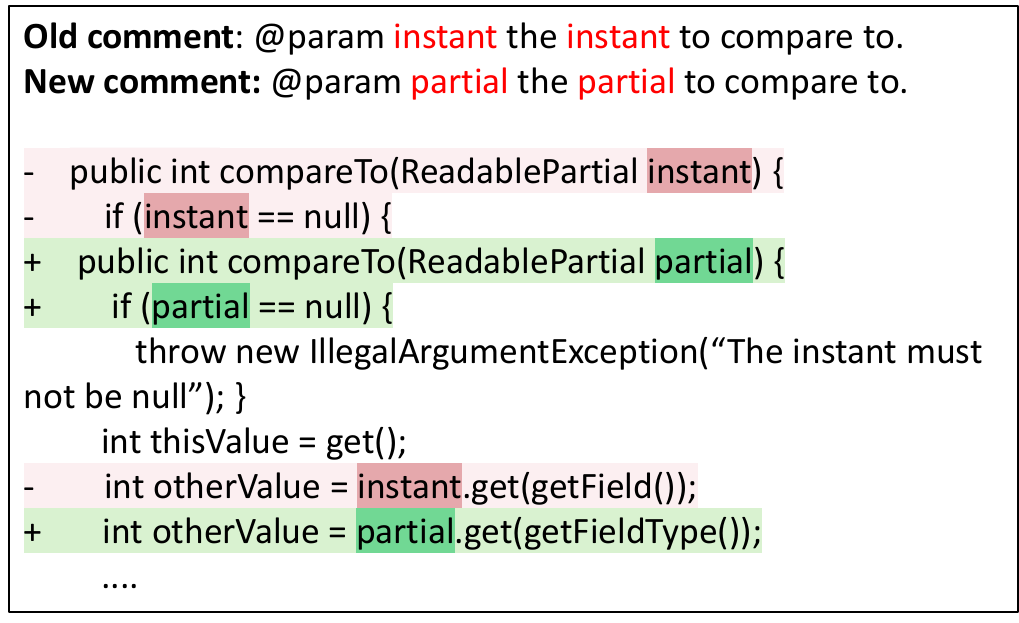}
        \label{fig:example2}
    }
    \hfill
    \subfloat[The example of summary inconsistency.]{%
        \includegraphics[width=0.81\linewidth]{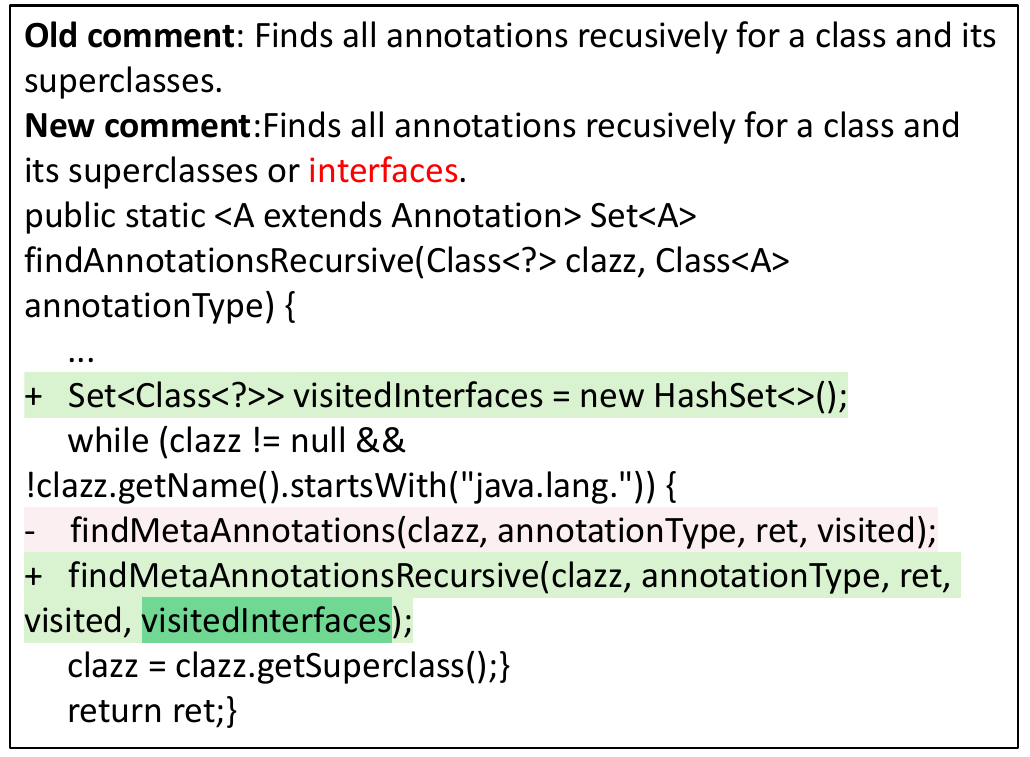}
        \label{fig:example3}
    }
    \caption{Three inconsistent cases in \jitdata, where old comment is inconsistent with the new code.}
    \label{fig:incon_example}
    \vspace{-10pt}
\end{figure}

Figures~\ref{fig:incon_example} demonstrate three distinct CCI cases. In the instance of return type inconsistency depicted in Figure~\ref{fig:example1}, the code modifies the return type from a list of FontIconSets to a collection of registered IconSets. Nonetheless, the original comment claims that the method supplies a list of registered FontIconSets. This results in a mismatch between the return type described by the comment and the actual code. Figure~\ref{fig:example2} illustrates an inconsistency in the method signature where the original comment and method parameter referred to `instant', but the revised code substituted it with `partial'. Figure~\ref{fig:example3} presents a case of summarization inconsistency, where the code introduces a set named visitedInterfaces and replaces the findMetaAnnotations call with findMetaAnnotationsRecursive, signaling a logical change.

\begin{figure}
    \centering
    \includegraphics[width=0.8\linewidth]{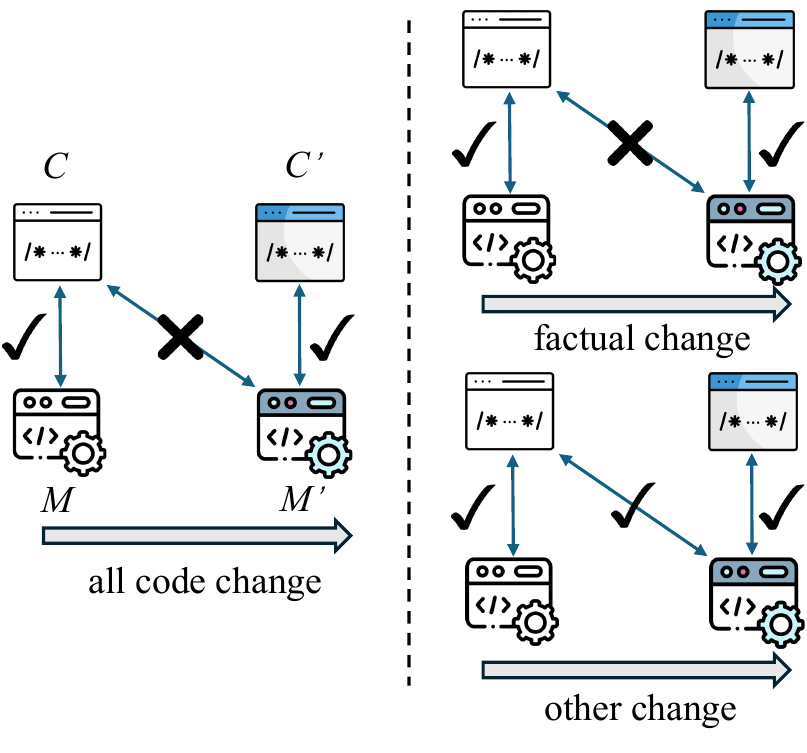}
    \caption{The presumptive scenario (left portion) and the actual scenario (right portion) that lead to CCI.}
    \label{fig:assumption}
    \vspace{-10pt}
\end{figure}

However, the data collection process of \jitdata can introduce labeling errors (i.e., false positives). Specifically, they extract the code-comment pair from the old and new versions ($(M$,$C$) ,($M{\prime}$,$C{\prime}$)) where $C \neq C{\prime}$. By assuming that the developer updated the comment because it would have otherwise become inconsistent as a result of code changes, they take $C$ to be inconsistent with $M{\prime}$~\cite{panthaplackel2021Deepa}, as the left part of Figure~\ref{fig:assumption} shows. While it is convenient for data collection, the assumptions are not always valid in practice~\cite{xu2023Data}. For example, if $C$ is updated because of typos or refactoring, despite it actually being consistent with $M{\prime}$, it would be assigned as an inconsistent case.

According to Definition~\ref{de:inconsistency}, we focus primarily on factual inconsistencies between code and comments, as depicted on the right side of Figure~\ref{fig:assumption}. Only some code changes can lead to a CCI occurrence. Consequently, it is crucial to determine what other types of changes can take place and their frequency, to assist in devising a method for their exclusion.

\subsection{Motivating Study}
This study aims to quantitatively investigate the prevalence of false positives in \jitdata and their taxonomy.
\subsubsection{Data sampling}
To conduct manual analysis, we randomly sample 200 positive cases each from @return, @param, and summary from \jitdata respectively (600 in total) based on a 95\% confidence level and 5\% confidence interval~\cite{boslaugh2012statistics}. For each sampled positive code-comment inconsistency case, we obtain the old comment, old code, new code, new comment, and differentiate between old and new code for further inspection. 

\subsubsection{Two-stage examination process}
\label{label_process}
\added{To ensure a high-quality manual analysis, we establish a multi-tiered annotation protocol. The process is initiated by two experienced co-authors (A1 and A2), each with over seven years of programming experience, who served as the primary annotators. They follow a two-step procedure: First, they independently analyze and label each sampled case. Second, for any individual case where their labels disagreed, they discuss the issue to reach a mutual consensus. For the rare and particularly challenging cases where an agreement could not be established through discussion, a third senior author, acting as an expert arbiter, is consulted to make the final decision. This multi-tiered process ensures that all disagreements are resolved rigorously, thereby guaranteeing the consistency and high quality of the final labeled dataset for subsequent analysis. The specific steps of the study are as follows:}

\textit{Step I}: A1 and A2 independently analyze all 600 sampled cases to decide inconsistencies. They then compare their results, discussing any disagreements until they reach a consensus. The level of agreement before discussion is indicated by a Cohen kappa of 0.76, reflecting a substantial level of agreement~\cite{sim2005kappa}.

\textit{Step II}: After receiving the false positive case, A1 randomly selects 30\% of them and reviews the difference between the code changes and the semantic meaning of the comments. A1 further drafts a list of categories of reasons that cause false positives. Then A1 and A2 follow the draft list to label the 30\% samples collaboratively to revise and refine the categories. Then A1 and A2 independently study the rest of the cases. The results in this phase have a Cohen's Kappa of 0.91, which is an almost perfect agreement.


\subsubsection{Study Result}\label{sec:study}

\begin{table}[tbp]
\centering
\caption{Preliminary study result.}
\label{tab:preliminary}
\scriptsize
\begin{tabular}{@{}lcccc@{}}
\toprule
                         & @return & @param & summary & Total \\ \midrule
Real Inconsistency       & 113     & 124    & 89      & 326   \\ \midrule
False Positive           & 87      & 76     & 111     & 274   \\
\hspace{10pt}-- add/delete information & 69      & 60     & 90      & 219   \\
\hspace{10pt}-- fix typo               & 4       & 3      & 9       & 16    \\
\hspace{10pt}-- change case            & 5       & 7      & 4       & 16    \\
\hspace{10pt}-- change lexical         & 9       & 6      & 8       & 23    \\ \bottomrule
\end{tabular}
\vspace{-10pt}
\end{table}


Table~\ref{tab:preliminary} presents the findings of the preliminary study. According to the analysis, 54.33\% (326 out of 600) of the instances are recognized as inconsistencies, indicating a true contradiction between the outdated comments and the updated code. The other 45.67\% (274 out of 600) are categorized as false positives. Among false positives, the primary causative factors are detailed as follows: Add/Delete information (219 instances, 80\%) is the leading contributor to false positives, reflecting changes where comments are elaborated or reduced for enhanced clarity. Typo correction (16 instances, 6\%) encompasses the rectification of errors such as misspellings in the comments. Change case (16 instances, 6\%) pertains to modifications in letter casing, including capitalizing or decapitalizing words, which are minor formatting adjustments. Change lexical (23 instances, 8\%) involves synonym substitution or rewording. Table~\ref{tab:fp_example} gives examples for each false positive categories.

\begin{table}[tbp]
\centering
\caption{Four false positive examples.}
\label{tab:fp_example}
\resizebox{\columnwidth}{!}{
  \begin{tabular}{@{}ll@{}}
    \toprule
    Type & Comment change \\ \midrule
    \multirow{2}{*}{add/delete information} & Removes the account \textcolor{blue}{with the primary key} from the database. \\
     & Removes the account from the database. \\
    \multirow{2}{*}{fix typo} & Returns the result of \textcolor{blue}{interpretting} the object as an instance of `Dial Region'. \\
     & Returns the result of \textcolor{blue}{interpreting} the object as an instance of `Dial Region'. \\
    \multirow{2}{*}{change case/stopwords} & Provides \textcolor{blue}{a} string representation of the property. \\
     & Provides \textcolor{blue}{the} string representation of the property. \\
    \multirow{2}{*}{change lexical} & \textcolor{blue}{Check} if specified address is allowed by current IPAccess rules. \\
     & \textcolor{blue}{Checks} if specified address is allowed by current IPAccess rules. \\
    \bottomrule
  \end{tabular}
}
\vspace{-10pt}
\end{table}

The study results reveal that many false positives stem from modifications aimed at improving comment quality or formatting rather than from any actual conflict with the code. The relatively small percentage of typo corrections, case changes, and lexical updates further support that most false positives arise from content enhancement or reduction. Additionally, the results emphasize the importance of carefully distinguishing real inconsistencies from false positives in the dataset construction process, as the latter can inflate results and misrepresent the tool’s performance.

\begin{tcolorbox}[boxsep=1pt,left=2pt,right=2pt,top=3pt,bottom=2pt,width=\columnwidth,colback=white!95!black,boxrule=1pt, colbacktitle=white!30!black,toptitle=2pt,bottomtitle=1pt,opacitybacktitle=0.4,fonttitle=\small,fontupper=\small]
\textbf{Summary:} 
\added{Our analysis of a representative sample from the JITDATA dataset suggests that a substantial portion of cases, approximately 45.67\%, are likely false positives according to our definition. Among the false positive samples, we identify four reasons: add/delete information, fix typo, change case, and change lexical. To evaluate the performance of the CCI task tool, it is essential to construct a new dataset.}
\end{tcolorbox}
\section{Dataset Construction -- \bench}
\label{dataset_construction}

\begin{figure*}
    \centering
    \includegraphics[width=0.85\textwidth]{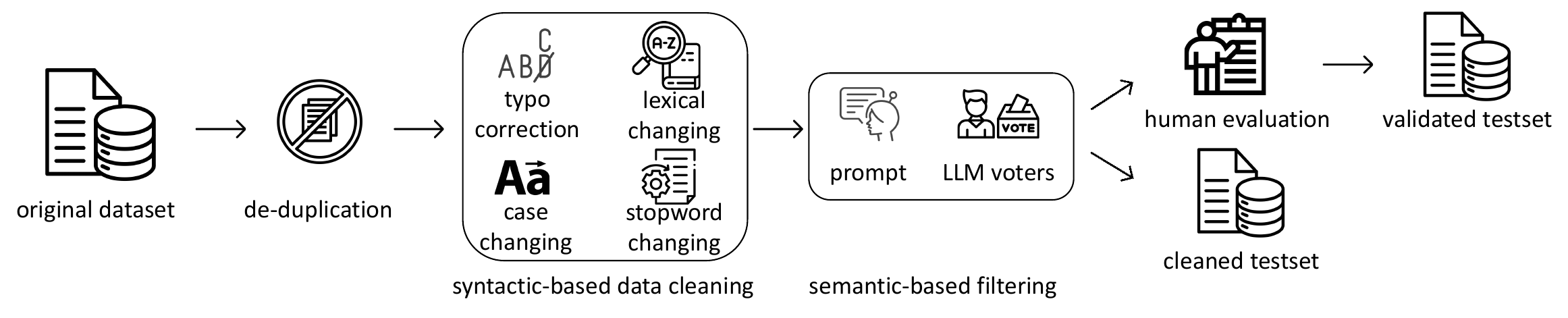}
    \caption{\bench construction process.}
    \label{fig:bench_construction}
    \vspace{-10pt}
\end{figure*}

In this section, we introduce a new dataset, \bench, containing an automated pipeline to acquire high-quality samples with accurate labels, as shown in Figure~\ref{fig:bench_construction}.

\subsubsection{Data source and data de-duplication}
\added{We choose the \jitdata as our data source due to its popularity in the past CCI-related studies~\cite{panthaplackel2021Deepa,rong2024code,dau2024DocChecker,steiner2022Code}.
This step aims to remove duplicates and cases with only formatting differences. To achieve this, for each case, we first normalize the codes and comments by stripping away characters such as tabs (``$\backslash$t'') and newlines (``$\backslash$n''). We consider cases as duplicates only if they share identical values for $M_{old}$, $M_{new}$, $C_{old}$, and $C_{new}$.  Labels are excluded from consideration due to potential labeling errors. For duplicate cases, we prioritize keeping examples with a true label. If all labels are false, we randomly retain one. 
The thorough data de-duplication prevents any vulnerable functions from leaking from the training set into the test set, and avoids label conflicts. After de-duplication of the initial 40,688 \jitdata instances, removing 337 duplicates, 40,351 unverified entries remain. Manually labeling this extensive dataset is impractical.}

\subsubsection{Syntactic-based data cleaning}
Inspired by the preliminary study results and Xu et al.~\cite{xu2023Data}, many false positives arise from syntactic changes in comments. To address this, we design our filter rules using a trial-and-error method. Initially, we create rules based on manually labeled false positives from Section~\ref{preliminarystudy}. We then manually review 20\% of the filtered results to further enhance the rules for better high precision as the upcoming next phase can partially address syntactic false positives. Ultimately, we establish the following rules to detect four types of syntactic changes:
\begin{itemize}
    \item \textbf{Typo correction.} If and only if one comment word is changed, the minimum edit distance is smaller than 4, and the word before editing is not in the old code.
    \item \textbf{Case changing.} If and only if the case of one or more words in the comment is changed, and the words before editing are not in the old code.
    \item \textbf{Stopword changing.} If one or more of the following words ([a, an, the, in, on, at, by]) are added or deleted. 
    \item \textbf{Lexical changing.} If and only if the root word after the lemmatization of the changed word is the same, and the word before editing is not in the old code.
\end{itemize}

\subsubsection{Semantic-based false positive filtering}
\added{Based on the results of the preliminary study, 80\% of the false positives are caused by add/delete information, which cannot be filtered out by the above syntactic-based method. This is because such changes require an understanding of the factual and logical meaning of the code and comment, a task we define as semantic analysis. Syntactic cleaning, by contrast, only addresses superficial textual patterns. We therefore choose to leverage the capabilities of LLMs to filter out these semantically-driven false positives, as LLMs have demonstrated effectiveness in comprehending and labeling code-related tasks~\cite{zheng2023judging}.}

\added{In this step, we employ a trio of popular and powerful LLMs to develop a voting mechanism: GPT4o~\cite{OpenAIGPT4o}, Claude3.5-Sonnet~\cite{anthropic2024claude}, and LLama3.1-405b~\cite{llama3modelcard}. These models are selected not merely for their widespread use, but for their state-of-the-art performance in code tasks~\cite{livecodebench}, their diverse training methodologies and architectures which reduce the likelihood of correlated errors, and their prevalent use in recent software engineering studies~\cite{gao2024search,zhong2025larger}.
The voting mechanism aggregates responses from multiple LLMs and selects the common one~\cite{aroraask}, maximizing the models' capabilities and enhancing judgment accuracy.
Specifically, if the two-thirds majority or all of the voters identify an inconsistency between a comment and the corresponding code, we choose to keep this case in our cleaned dataset. Conversely, if fewer than two-thirds register an inconsistency, we discard the data.}

\added{Appendix A provides a detailed prompt used for this semantic filtering. In addition to this mechanism, we apply a dynamic 4-shot in-context learning (ICL) strategy. This is crucial, as it aids the LLM voters in comprehending the task at hand more precisely~\cite{gao2023makes}, thereby allowing them to concentrate on specific semantic discrepancies or factual contradictions.
The four examples are specifically chosen to cover the different types: one example for a consistent case, and one for each of the three inconsistency types (return type, method signature, and application logic). Note that all filtering query use the same four examples.
After the above process, the statistics of \bench are illustrated in Table~\ref{tab:dataset_statistics}.}

\subsubsection{Validated test set construction}
Following previous studies~\cite{panthaplackel2021Deepa}, we created a validated subset within the test set to better assess the effectiveness of methods. First, we involve cases that were unanimously deemed inconsistent by all three LLM voters, indicating a strong likelihood of inconsistency. Then, two annotators manually verify the label corrections, following a process similar to that in Section~\ref{label_process}. The annotators achieved a Cohen's Kappa of 0.95, indicating almost perfect agreement and highlighting our annotation reliability. Ultimately, we get 300 validated test cases.

\begin{table}[!t]
\centering
\caption{Statistics of the \jitdata and \bench in different inconsistency and set types.}
\scriptsize
\label{tab:dataset_statistics}
\begin{tabular}{@{}clrrrr@{}}
\toprule
\multicolumn{1}{l}{}      & Type    & Train & Valid & Test & Total \\ \midrule
\multirow{4}{*}{\jitdata} & return  & 15,950 & 1,790  & 1,840 & 19,580 \\
                          & param   & 8,640  & 932   & 1,038 & 10,610 \\
                          & Summary & 8,398  & 1,034  & 1,066 & 10,498 \\
                          & Full    & 32,988 & 3,756  & 3,944 & 40,688 \\ \midrule
\multirow{4}{*}{\bench}  & return  & 9,188  & 1,092  & 1,088 & 11,368 \\
                          & param   & 6,004  & 620   & 646  & 7,270  \\
                          & Summary & 2,970  & 356   & 396  & 3,722  \\
                          & Full    & 18,162 & 2,068  & 2,130 & 22,360 \\ \bottomrule
\end{tabular}
\vspace{-10pt}
\end{table}
\vspace{-5pt}
\section{Methodology}
\label{method}

\subsection{Overview}
\begin{figure}
    \centering
    \includegraphics[width=1\linewidth]{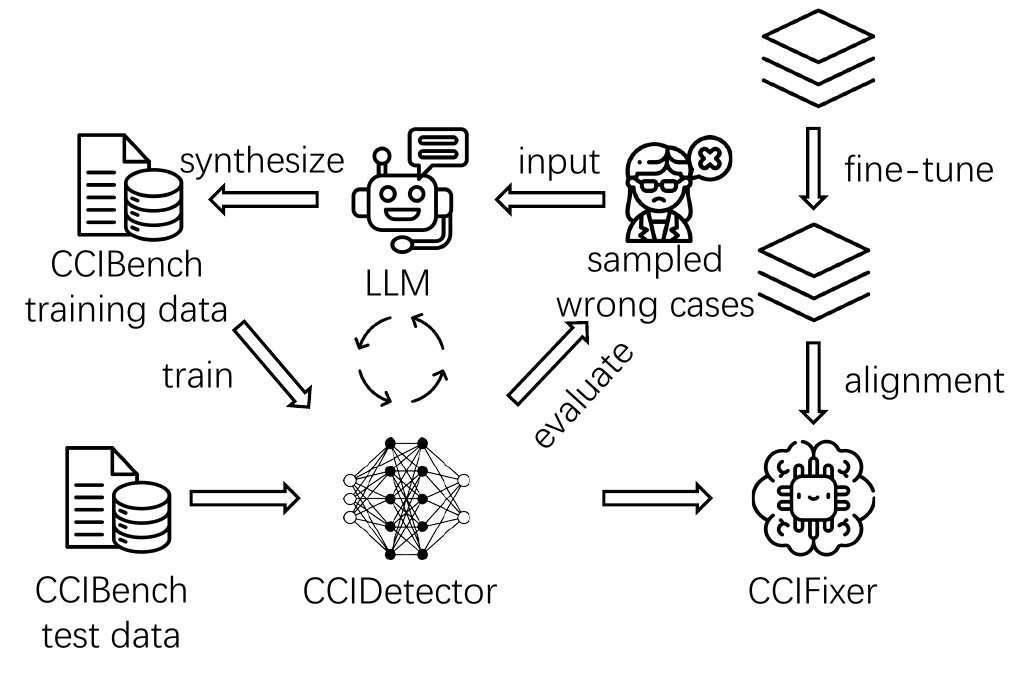}
    \caption{The framework of \name.}
    \label{fig:framework}
    \vspace{-10pt}
\end{figure}
In this section, we introduce \name, an end-to-end framework to tackle the CCI problem effectively.
This structure of \name are shown in Figure~\ref{fig:framework}, which consists of two main components: the \detector to identify inconsistent cases, and the \fixer, which corrects these inconsistencies. In the first stage, \detector obtains the code comments and related code changes from the training data of \bench. After encoding with various techniques, it undergoes a semantic similarity classifier. To continuously combine knowledge from failure examples, we propose an LLM-based enhancement technique to iteratively synthesize challenging data that are mistakenly classified. In the second stage, \fixer is implemented by using the parameter-efficient fine-tuning (PEFT) methods to fine-tune and align the pre-trained LLM, using the inconsistent cases in training data in \bench.


\begin{figure} [tbp]
    \centering
    \includegraphics[width=0.95\linewidth]{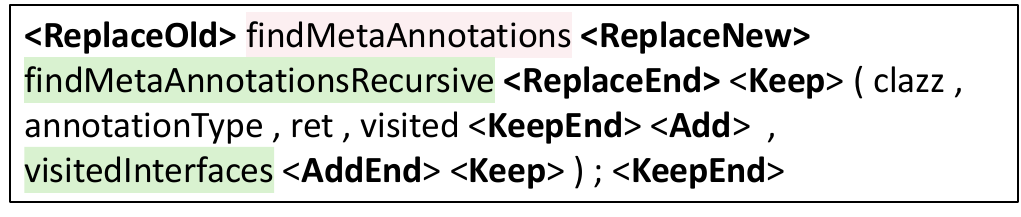}
    \caption{Code diff representation related to Figure~\ref{fig:example3}, removed tokens in red and added tokens in green.
    }
    \label{fig:code_edit}
    \vspace{-10pt}
\end{figure}
\subsection{\detector}
\subsubsection{Code diff representation}
In the domain of code diff representation, employing semantically identical yet formally distinct representations concurrently as input has yielded remarkable performance results~\cite{panthaplackel2021Deepa,leclair2019neural}. 
In this method, we utilize difflib~\cite{python_difflib} to extract code diffs based on the study by Panthaplackel et al.~\cite{panthaplackel2021Deepa}, as previous work indicates that this diff structure outperforms the token-level diff structure~\cite{panthaplackel2020Learning}. Specifically, we employ four edit actions: \textit{add}, \textit{del}, \textit{rep}, and \textit{keep}. The actions \textit{add}, \textit{del}, and \textit{keep} are formatted as $\langle Action\rangle [tokens] \langle EndAction\rangle$. The \textit{replace} action, which incorporates content from both the old and new versions, is structured as $\langle ReplaceOld\rangle [tokens] \langle ReplaceNew\rangle [tokens] \langle ReplaceEnd\rangle$. We extract the diff between $M_{old}$ and $M_{new}$ using the above diff lexicon. Figure~\ref{fig:code_edit} shows the $M_{edit}$ corresponding to code diff in Figure~\ref{fig:example3}. 

\subsubsection{Encoding}
For raw code and comment, we use the encoder-only Unixcoder~\cite{guo2022unixcoder} to embed the raw code and comment into fixed-size vectors. Unixcoder is a unified pre-trained model for programming languages that are based on a multi-layer transformer and utilizes mask attention matrices~\cite{dong2019unified} with prefix adapters to control the access to context for each token and has been widely used in code-related tasks~\cite{zhang2024automatic,guo2024exploring}. Note that the training process is end-to-end where the parameters of the encoder layers are adjusted to enhance the representation vectors, allowing these vectors to better adapt to the CCI detection task.

For the code diff representation, following the previous work~\cite{xu2023Data,xu2024Code}, we use Bi-GRU layers to encode the diff representation. Such encoder specifically models token representation in code diffs by considering the current token \( x_t \), the previous hidden state \( h_{t-1} \), and contextual information from both directions, which enhances the model's grasp of contextual dependencies. Each code diff is tokenized and processed through a Bi-GRU. At each step \( t \), the current token and previous hidden state update the current hidden state $h_t = \text{GRU}(x_t, h_{t-1})$.
The Bi-GRU captures both forward and backward information. At each time step, the output combines the forward \( \overrightarrow{h_t} \) and backward \( \overleftarrow{h_t} \) hidden states $h_t = W_f \cdot \overrightarrow{h_t} + W_b \cdot \overleftarrow{h_t} + b$
where \( W_f \) and \( W_b \) are the respective weight matrices, and \( b \) is a bias. Given that code diffs involve long-range dependencies, we also use Multi-Head Self-Attention for capturing global context. This mechanism evaluates token interrelationships using Query (\( Q \)), Key (\( K \)), and Value (\( V \)) vectors: $\text{Attention}(Q, K, V) = \text{softmax}\left(\frac{Q K^\top}{\sqrt{d_k}}\right)V$
where \( d_k \) scales the softmax. Multi-Head Attention uses several heads to focus on various code diff aspects: $\text{MultiHead}(Q, K, V) = \text{Concat}(\text{head}_1, \dots, \text{head}_h) W^O$. Each attention head is: $
\text{head}_i = \text{Attention}(Q W_i^Q, K W_i^K, V W_i^V)$ This multi-head approach enhances the encoder's ability to capture diverse representations.

\subsubsection{Semantic similarity classification}
After obtaining representations for $C_{old}$ and $M_{new}$ ($M_{diff}$), a classifier head is devised to identify CCI. Instead of using a basic classifier with fully connected layers, we introduce a similarity-based classification model as comments should align with the semantic meaning of adjacent code. The classifier concatenates the embeddings of the comment $C_{old}$ and its corresponding code snippet $M$. The loss function, detailed in Equation~\ref{loss}, combines binary cross-entropy (BCE) loss and a similarity-based term. The BCE loss is standard in machine learning, while the similarity-based term assesses semantic alignment between comments and code. A weighting parameter $\lambda$ balances the influence of each term, with an additional constraint ensuring the total loss remains positive.

\begin{align}
\label{loss}
   \mathcal{L} = - &(\frac{1}{N} \sum_{i=1}^N \left[ y_i \log(p_i) + (1 - y_i) \log(1 - p_i) \right] \nonumber \\ &+ \lambda\frac{1}{N}\sum_{i=1}^{N}\frac{c_i \cdot m_i}{\|c_i\| \|m_i\|} - \lambda)  
\end{align}

\subsubsection{Iterative enhancement}
As data enhancement becomes popular~\cite{wei2019eda,shorten2021text,bouzenia2023say}, we aim to boost \detector's performance from the perspective of training data. A straightforward approach is to use a prompted LLM to augment the entire dataset~\cite{dai2023auggpt}, but this has restrictions in considering the model's varying performance across different data. For instance, while \detector may handle the majority of the dataset, it may struggle with a few challenging cases. Rather than duplicating simple cases, it is more beneficial to generate challenging and difficult examples.

\added{Inspired by using LLMs for data augmentation~\cite{lee2024llm2llm} and the Self-Instruct framework~\cite{wang2022self}, we introduce an iterative methodology to improve the training dataset, successfully mitigating challenges associated with certain error-prone instances of the \detector. Specifically, in each training iteration, we identify cases where the model predicts incorrect classifications. Then, to ensure diversity and avoid over-concentration on a narrow subset of errors, we sample a fraction of these incorrect cases (e.g., with a default sampling rate of 0.1). We then utilize a teacher LLM to generate new samples similar to these selected difficult ones.}

Algorithm 1 shows the iterative enhancement procedure. The input contains the original cleaned dataset $D^0$, original detection model and the teacher LLM. The process begins with training the base \detector model on the initial dataset $D^0$, establishing a foundational detection performance. 

\added{Afterward, the model is evaluated on the current training set, and the evaluation results are analyzed to pinpoint instances of misclassification errors.
Instead of feeding all problematic cases into the LLM, which could bias the augmentation towards the most frequent error types, we randomly sample a subset of these instances based on a predefined sampling rate (defaulting to 0.1). }
These selected cases, which highlight the model’s limitations, are then fed into the LLM. By carefully designing prompts for the LLM, synthetic examples are generated. These examples maintain conceptual consistency with the original cases to ensure task relevance while introducing semantic diversity to improve the model’s robustness.
The synthesized data is subsequently merged with the existing training dataset to produce an updated training set $D^{i+1}$, which is then used for the next round of training. This iterative process continues, enhancing the model’s ability to handle challenging cases, until a termination criterion is met, such as achieving a desired performance level or observing convergence in model improvement. 

Similar to \cite{lee2024llm2llm}, we only use cases from the original data $D^0$ to create additional data to avoid degradation from multiple iterations. While LLMs can produce high-quality cases, errors may still occur. Further expanding these flawed examples can spread errors and compromise dataset quality. The details of designed prompts and examples are shown in Appendix A.



\begin{algorithm}
\caption{Iterative enhancement procedure}
\KwIn{The original cleaned dataset $D^0$, original CCIDetection model $M_{detect}$, and LLM $T$}
\KwOut{Enhanced training dataset $D^i$}
$i \gets 0$ \;
\While{$true$}{
    $M_{detect}^i \gets \text{Train}(M_{detect}^0,D^i)$\;
    $E_{detect}^i \gets \text{Evaluate}(M_{detect}^i,D^0)$ \;
    \If{$\text{stop condition}$}{
        \Return $D^i$ \;
    }
    $D_{synthetic}^i \gets \text{Synthesize}(E_{detect}^i,T)$ \;
    $D^{i+1} \gets D^{i} + D_{synthetic}^i$
    
}
\end{algorithm}

\vspace{-10pt}
\subsection{CCIFixer}

\subsubsection{Fine-tuning}
\added{
Using our cleaned dataset, we fine-tune the base LLM to develop our inconsistency fixing model, \fixer.
We employ LoRA~\cite{hu2021lora} for our fine-tuning process, an efficient parameter-efficient fine-tuning (PEFT) scheme.
This approach is advantageous as it significantly reduces the resource demands for fine-tuning tasks while achieving results comparable to full-parameter fine-tuning.
It accomplishes this by utilizing a small fraction (e.g., as little as 1.5\%) of the original model's parameters for the fine-tuning process.
The underlying principle of LoRA is based on the hypothesis that the change in weights during model adaptation has a low intrinsic rank.
Thus, the update to a weight matrix $\Delta W$ can be decomposed into a product of two low-rank matrices.
Mathematically, for a pre-trained weight matrix $W_0 \in \mathbb{R}^{d \times k}$, the fine-tuned matrix $W'$ is expressed as:
$W' = W_0 + \Delta W = W_0 + BA$.
Here, $\Delta W$ represents the change after fine-tuning, which is the product $BA$, where $B \in \mathbb{R}^{d \times r}$ and $A \in \mathbb{R}^{r \times k}$.
The dimensions $d$ and $k$ correspond to the dimensions of the original weight matrix $W_0$, while $r$ is the rank, satisfying $r \ll \min(d, k)$.
During the fine-tuning stage, $W_0$ is kept frozen (i.e., not subject to gradient updates); only the matrices $A$ and $B$ contain trainable parameters and are updated.
Importantly, for an input $x$, both $W_0$ and $\Delta W = BA$ process $x$, and their respective output vectors, $W_0x$ and $BAx$, are summed element-wise: $h = W_0x + BAx = (W_0 + BA)x$.
The significant reduction in trainable parameters (from $dk$ for $W_0$ to $r(d+k)$ for $A$ and $B$) makes it feasible to fine-tune large models with minimal parameter overhead.
}

\subsubsection{Alignment}
Although fine-tuning large models on task-specific data can improve performance, the creative capabilities of these models in a CCI fix task may result in comments that are semantically correct but significantly different from the ground truth when evaluated with text similarity metrics. Therefore, we employ the process of alignment to guide the model in learning the overall programming preferences in \bench.

Conventional alignment methods require paired data, such as preferred and dispreferred data. In this specific code comment inconsistent fix task, it is not possible to acquire the dispreferred data naturally. Therefore, we need an alignment algorithm that does not rely on paired data.
Inspired by Ethayarajh et al.~\cite{ethayarajh2024kto}, we introduced the Kahneman-Tversky Optimization (KTO) algorithm to our alignment process. Unlike traditional methods like RLHF or DPO~\cite{rafailov2024direct} that rely on ranked comparisons, KTO models human preferences through a utility-based objective derived from the perceived gains and losses in prospect theory. The utility function $U(x)$ is defined relative to a reference point $r$, with risk sensitivity parameters $\alpha, \beta$ and a loss aversion factor $\lambda$, capturing human behavior toward gains and losses. The model optimization directly minimizes the Human Aligned Loss Objective (HALO), $L_{\text{HALO}} = \mathbb{E}_{x \sim p_\theta(x)} \left[ -U(x) \right]$, where $p_\theta(x)$ represents the model’s output distribution. Specifically, the default KTO optimization task is:
\begin{equation}
    L_{\text{KTO}}(\pi_\theta, \pi_{\text{ref}}) = \mathbb{E}_{x, y \sim \mathcal{D}} \left[ \lambda_y - v(x, y) \right], 
\end{equation}

where 
\begin{equation}
    r_\theta(x, y) = \log \frac{\pi_\theta(y|x)}{\pi_{\text{ref}}(y|x)}
\end{equation}
\begin{equation}
    z_0 = \text{KL}(\pi_\theta(y'|x) \| \pi_{\text{ref}}(y'|x))
\end{equation}
\begin{equation}
    v(x, y) =
    \begin{cases} 
    \lambda_D \sigma \left( \beta \big( r_\theta(x, y) - z_0 \big) \right) & \text{if } y \sim y_{\text{desirable}} \mid x \\
    \lambda_U \sigma \left( \beta \big( z_0 - r_\theta(x, y) \big) \right) & \text{if } y \sim y_{\text{undesirable}} \mid x
    \end{cases}
\end{equation}

$r_\theta(x, y)$ is a reward function that measures the deviation of the model's predictions from a reference distribution. A baseline $z_0$ normalized the reward, accounting for the overall divergence. The utility $v(x, y)$ is then calculated differently for desirable and undesirable outputs, scaled by parameters $\lambda_D, \lambda_U$, and adjusted by a sensitivity factor $\beta$.

\section{Experiment Setup}
\label{experiment}
\subsection{Baselines}

\added{To fairly evaluate \name's performance, we carefully choose open-source baselines through a brief review of related papers from SE venues in the past decade. Furthermore, we take general-purpose LLMs and code-based LLMs into consideration, as they have been demonstrated excellent performance in SE tasks~\cite{tian2024large,li2024exploring,geng2024Large}.}

\subsubsection{Learning-based method}
We compare our approach against a variety of learning-based baselines. These include methods that leverage large pre-trained models like \textbf{CodeBERT BOW}~\cite{feng2020codebert} and the use of \textbf{Bert} and \textbf{Longformer}~\cite{steiner2022Code} to better handle long sequences. Other methods focus on explicitly modeling code structures and changes. For instance, \textbf{DeepJIT}~\cite{panthaplackel2021Deepa} explores sequential (SEQ), graph-based (GRAPH), and hybrid encoders (HYBRID). \textbf{OCD}~\cite{liu2023JustInTime} represents code edits as triple sequences processed by Bi-LSTM and Co-Attention layers, while \textbf{AdvOC}~\cite{xu2023Data} employs a Gated Graph Neural Network on edit trees within an adversarial learning framework. \textbf{DocChecker}~\cite{dau2024DocChecker} utilizes the UniXcoder model~\cite{guo2022unixcoder} for just-in-time inconsistency detection.

\subsubsection{General-purpose LLMs} 
\added{In this study, we select three widely-used general-purpose LLMs as baselines: \textbf{GPT-3.5}~\cite{gpt-3.5}, \textbf{GPT-4o}~\cite{OpenAIGPT4o}, and \textbf{LLaMA-3.1}~\cite{llama3modelcard}. These models are chosen due to their extensive adoption and proven effectiveness in SE tasks~\cite{tian2024large,li2024exploring,geng2024Large}. Their robust performance in understanding and generating code-related text makes them suitable benchmarks for evaluating our proposed framework.}


\subsubsection{Code-based LLMs}
Building on the success of LLMs in natural language processing, researchers have also developed code-oriented LLMs to enhance code comprehension and task generation. In this study, we choose \textbf{Deepseek-coder}\cite{zhu2024deepseek} to deal with the CCI task. Furthermore, \textbf{C4RLLaMA}~\cite{rong2024code} is a fine-tuned large language model based on the CodeLLaMA and shows promising results for CCI tasks.

\vspace{-5pt}
\subsection{Eavaluation metrics}

\subsubsection{Evaluation Metrics for CCIDetector}
The detection of code comment inconsistency is a binary classification problem~\cite{liu2023JustInTime}. Based on the methodology outlined in previous work~\cite{panthaplackel2021Deepa}, we evaluated the performance of the related approaches using Precision, Recall, F1-score, and Accuracy metrics. 

\subsubsection{Evaluation Metrics for CCIFixer}
To rigorously evaluate the performance of CCIFixer, our methodology employs metrics targeting both text generation and text editing tasks, in line with the previous work~\cite{panthaplackel2021Deepa}. For assessing generation quality, we utilize BLEU-4~\cite{papineni2002bleu} to measure n-gram precision against reference comments, providing an indication of surface-level fidelity. This is complemented by METEOR~\cite{banerjee2005meteor}, which offers a more semantically-aware evaluation by considering synonyms and stemming. To assess the model's effectiveness from a text-editing perspective, we adopt SARI~\cite{xu-etal-2016-optimizing} to specifically quantify the quality of the performed edit operations (i.e., additions, deletions, and retentions), and GLEU~\cite{napoles2015ground} to measure the similarity between the generated output and the reference, a standard approach for grammatical error correction tasks. Together, these four metrics provide a comprehensive, multi-faceted evaluation of the system's output quality.

\subsection{Implementation and Environment}
For hyperparameters in our proposed approach, we set the train epoch number as 10, the loss function weight $\lambda$ as 1, the default iteration number of enhancement is 10 for CCIDetector. In CCIFixer, the hyperparameters in fine-tuning stage and alignment stage are displayed in Table~\ref{tab:hyperparameters}.

We opted for GPT4o as the teacher LLM in \detector, primarily because of its advanced abilities in understanding and generating code. Meanwhile, for the \fixer base model, we selected Qwen2.5-Coder-14B. This decision was influenced by the necessity to balance device resource limitations with the requirement for proficient code and prompt comprehension ability.

\begin{table}[tbp]
\centering
\scriptsize
\caption{Fine-tuning hyperparameters with LoRA.}
\label{tab:hyperparameters}
\begin{tabular}{cccccccc}
\toprule
stage & epoch & bs & lr & max\_len & $lora_r$ & $lora_a$ & $lora_d$ \\ \midrule
fine-tune& 10 &   16   & 1e-5   & 2048   & 8 &  32          &  0.05 \\
alignment & 5& 32 & 1e-5 & 2048 & 8 & 32 & 0.05 \\
\bottomrule
\end{tabular}
\end{table}

\added{For conventional baseline approaches, we reproduced them based on the replication packages released by the authors. Note that for C4RLLaMA, we re-implemented and fine-tuned using the exact same Qwen2.5-Coder-14B model as our backbone. This crucial step ensures that any observed performance differences between our method and the baseline are attributable to the proposed methodological advancements, rather than a disparity in the underlying power of the base models. We repeat five times and report the median results to avoid potential random bias. We ran them on a server with Intel Xeon Gold 6248R CPU, 512GB physical memory, and 2x NVIDIA A100 40G GPU. The OS version is Ubuntu 18.04.}

For LLMs, we called their official APIs to infer. We adapt 4-shot in-context learning for prompt techniques in the data cleaning process. We use the prompt in Appendix A for the iterative enhancement in the detection phase. As a baseline, we also retrieve 4 cases with the highest similarity using BM25~\cite{robertson2009probabilistic} algorithms to detect CCI. For all experiments related to LLM inference, we set the temperature to 0 so that LLM would generate the same output for the same query to ensure reproducibility. 


\vspace{-10pt}
\section{Experimental Results}
\label{experimentresult}

\subsection{Research Questions}
In this study, we investigate the following research questions (RQs) and provide a case study to understand the fixing ability of \fixer:

\begin{itemize}
    \item RQ1: How effective is \detector in CCI detection?
    \item RQ2: How do different design choices enhance the performance of \detector?
    \item RQ3: How effective is \fixer in fixing the inconsistency of the existing code comments?
    \item RQ4: How do LLM backbones and tuning process enhance the
performance of \fixer?
    \item \added{RQ5: How effective and efficient is \name at end-to-end CCI detection and repair compared to existing method?}
\end{itemize}

\subsection{RQ1: How effective is \detector in CCI detection?}

\subsubsection{Motivation}
The purpose of this RQ is two-fold.
First, we want to evaluate the performance of existing approaches on our new proposed clean dataset \bench.
Second, we want to investigate whether \detector outperforms existing baselines in detecting CCI.
\subsubsection{Approach}
To achieve our objective, we reimplement all the baseline methods. We apply these implementations to both the \jitdata~\cite{panthaplackel2021Deepa} and \bench. Each of these datasets provides a validated test set consisting of 300 cases. However, it is important to note that the process of creating the validated dataset differs, leading to their differing cases.

\begin{table*}[tbp]
\centering
\caption{Just-in-time CCI detection performance on DeepJIT and \bench.}
\scriptsize
\label{tab:detection_result}
\begin{tabular}{@{}cccccccccc@{}}
\toprule
\multirow{2}{*}{Dataset}                   & \multirow{2}{*}{Model}           & \multicolumn{4}{c}{Full Test Set}        & \multicolumn{4}{c}{Validated Test Set}    \\
                           &                 & Accuracy & Precision & Recall & F1-score & Accuracy & Precision & Recall & F1-score \\ \midrule
\multirow{18}{*}{\jitdata}  & CodeBERT BOW    & 72.04    & 68.87     & 73.24  & 70.99    & 75.33    & 77.14     & 72.00  & 74.48    \\
                           & Seq             & 79.46    & 83.78     & 73.07  & 78.06    & 80.67    & 81.94     & 78.67  & 80.27    \\
                           & Graph           & 79.74    & 79.94     & 79.41  & 79.67    & 82.00    & 82.43     & 81.33  & 81.88    \\
                           & Hybrid          & 78.22    & 80.23     & 74.89  & 77.47    & 85.21    & 83.45     & 86.43  & 84.91    \\
                           & Seq+features    & 81.56    & \textbf{87.75}     & 73.38  & 79.92    & 86.67    & 87.67     & 85.33  & 86.48    \\
                           & Graph+features  & 82.48    & 84.05     & 80.17  & 82.06    & 86.33    & 86.09     & 86.67  & 86.38    \\
                           & Hybrid+features & 82.15    & 82.85     & 81.08  & 81.96    & \textbf{89.00}    & \textbf{89.26}     & \textbf{88.67}  & \textbf{88.96}    \\
                           & DocCheaker      & 72.97    & 71.25     & 73.17  & 72.20    & 76.33    & 78.01     & 73.33  & 75.60    \\
                           & Bert            & 77.67    & 76.18     & 78.21  & 77.18    & 80.05    & 81.94     & 78.67  & 80.27    \\
                           & Longformer      & 78.75    & 74.98     & 79.82  & 77.32    & 81.43    & 83.33     & 80.00  & 81.63    \\
                           & OCD             & 76.50    & 82.92     & 66.75  & 73.96    & 74.67    & 75.34     & 73.33  & 74.32    \\
                           & AdvOC           & 80.28    & 84.37     & 75.28  & 79.57    & 79.82    & 83.33     & 76.67  & 79.86    \\
                           & C4RLLaMA        & 86.24    & 86.20     & 84.30  & 85.24    & 87.82    & 89.67     & 85.33  & 87.45    \\
                           & GPT3.5          & 53.13    & 67.21     & 46.28  & 54.82    & 58.33    & 69.40     & 52.67  & 59.89    \\
                           & GPT4o            & 77.74    & 79.25     & 75.33  & 77.24    & 75.67    & 75.16     & 76.67  & 75.91    \\
                           & LLaMA3.1        & 77.89    & 68.24     & \textbf{88.75}  & 77.16    & 78.00    & 75.61     & 82.67  & 78.98    \\
                           & Deepseek-coder  & 65.98    & 74.33     & 60.24  & 66.55    & 68.00    & 68.24     & 67.33  & 67.78    \\
                           & \cellcolor{gray!20} \detector     & \cellcolor{gray!20}\textbf{86.48}    & \cellcolor{gray!20}87.24     & \cellcolor{gray!20}85.17  & \cellcolor{gray!20}\textbf{86.19}    & \cellcolor{gray!20}88.74    & \cellcolor{gray!20}87.67     & \cellcolor{gray!20}88.67  & \cellcolor{gray!20}88.17    \\ \midrule
\multirow{18}{*}{\bench} & CodeBERT BOW    & 73.04    & 71.20     & 73.64  & 72.40    & 76.00    & 77.86     & 72.67  & 75.18    \\
                           & Seq             & 82.44    & 84.38     & 79.62  & 81.93    & 85.00    & 85.23     & 84.67  & 84.95    \\
                           & Graph           & 84.13    & 84.92     & 83.00  & 83.95    & 84.33    & 85.03     & 83.33  & 84.17    \\
                           & Hybrid          & 80.65    & 78.37     & 84.69  & 81.41    & 83.67    & 79.53     & 90.67  & 84.74    \\
                           & Seq+features    & 87.51    & 89.67     & 84.79  & 87.16    & 89.67    & 88.88     & 90.67  & 89.77    \\
                           & Graph+features  & 88.22    & \textbf{90.46}     & 85.44  & 87.88    & \textbf{92.00}    & 91.45     & \textbf{92.67}  & 92.06    \\
                           & Hybrid+features & 87.32    & 88.41     & 85.92  & 87.15    & 89.67    & 89.40     & 90.00  & 89.70    \\
                           & DocCheaker      & 76.46    & 75.13     & 77.27  & 76.18    & 76.67    & 75.00     & 80.00  & 77.42    \\
                           & Bert            & 82.04    & 82.03     & 81.68  & 81.85    & 83.00    & 80.00     & 88.00  & 83.81    \\
                           & Longformer      & 82.47    & 80.82     & 83.35  & 82.07    & 84.00    & 80.72     & 89.33  & 84.81    \\
                           & OCD             & 77.86    & 84.68     & 70.24  & 76.79    & 76.67    & 77.86     & 74.67  & 76.23    \\
                           & AdvOC           & 84.10    & 87.25     & 79.31  & 83.09    & 84.28    & 86.82     & 80.64  & 83.62    \\
                           & C4RLLaMA        & 89.16    & 88.24     & 89.10  & 88.67    & 87.82    & \textbf{92.67}     & 88.67  & 90.63    \\
                           & GPT3.5          & 64.41    & 70.72     & 49.20  & 58.03    & 63.00    & 69.70     & 46.00  & 55.42    \\
                           & GPT4o           & 83.19    & 84.76     & 80.94  & 82.81    & 82.67    & 88.89     & 74.67  & 81.16    \\
                           & LLaMA3.1        & 78.87    & 72.27     & 93.71  & 81.61    & 80.67    & 75.56     & 90.67  & 82.43    \\
                           & Deepseek-coder  & 74.27    & 80.45     & 64.13  & 71.37    & 77.67    & 84.30     & 68.00  & 75.28    \\
                           & \cellcolor{gray!20} \detector     & \cellcolor{gray!20} \textbf{89.92}    & \cellcolor{gray!20} 89.84     & \cellcolor{gray!20} \textbf{89.25}  & \cellcolor{gray!20} \textbf{89.54}    & \cellcolor{gray!20} 90.95    & \cellcolor{gray!20} \textbf{92.67}     & \cellcolor{gray!20} 91.74  & \cellcolor{gray!20} \textbf{92.20}    \\ \bottomrule
\end{tabular}
\vspace{-15pt}
\end{table*}

\subsubsection{Result}

Table~\ref{tab:detection_result} presents the just-in-time code comment inconsistency detection results on two datasets with four metrics: accuracy, precision, recall, and F1 score. 


The \detector method outperforms all other techniques in \bench, establishing itself as the most proficient strategy for just-in-time CCI detection. Within the validated test dataset, \detector secures an impressive F1 score of 92.20, considerably exceeding its nearest competitor, C4RLLaMA, which achieves an F1 score of 90.63 by a margin of 1.73\%. This consistent superiority underscores \detector’s robustness in managing high-confidence samples, maintaining a balance between precision and recall, and effectively generalizing across datasets with different degrees of complexity.

An interesting observation is that the performance improvement gained by adding features is significantly greater when using the \bench dataset compared to \jitdata. For example, on DeepJIT’s Full Test Set, Graph+features improves the F1-score from 79.67 to 82.06, an increase of 2.39 points. However, in \bench, the same feature enhancement yields a much larger improvement, with the F1 score rising from 83.95 to 87.88, a notable increase of 3.93 points. Similarly, for the Hybrid model, adding features in \jitdata from 77.47 to 81.96 (4.49 points), while in \bench, the improvement jumps from 81.41 to 87.15 (5.74 points). 

Although LLMs have shown exceptional performance in many natural language processing tasks, their effectiveness in CCI detection appears to lag behind most baseline methods. As observed in the results, models like GPT-3.5 and deepseek-coder achieve significantly lower performance metrics compared to tailored baseline methods. For example, on the validated test set in \bench, GPT-3.5 records a recall of only 46.00 and an F1 score of 55.42. Similarly, on the \jitdata dataset, the F1 score of deepseek coder (67.78) is outperformed by even traditional approaches such as CodeBERT Bow, which achieves 74.48. 

\begin{tcolorbox}[boxsep=1pt,left=2pt,right=2pt,top=3pt,bottom=2pt,width=\columnwidth,colback=white!95!black,boxrule=1pt, colbacktitle=white!30!black,toptitle=2pt,bottomtitle=1pt,opacitybacktitle=0.4,fonttitle=\small,fontupper=\small]
\textbf{RQ1 Result:} The \detector method exhibits better results across both datasets, thereby establishing it as the foremost technique for just-in-time CCI detection by 1.73\%.
\end{tcolorbox}

\vspace{-10pt}
\subsection{RQ2: How do different design choices enhance the performance of \detector? }

\begin{figure}[tbp]
    \centering
    \includegraphics[width=0.8\linewidth]{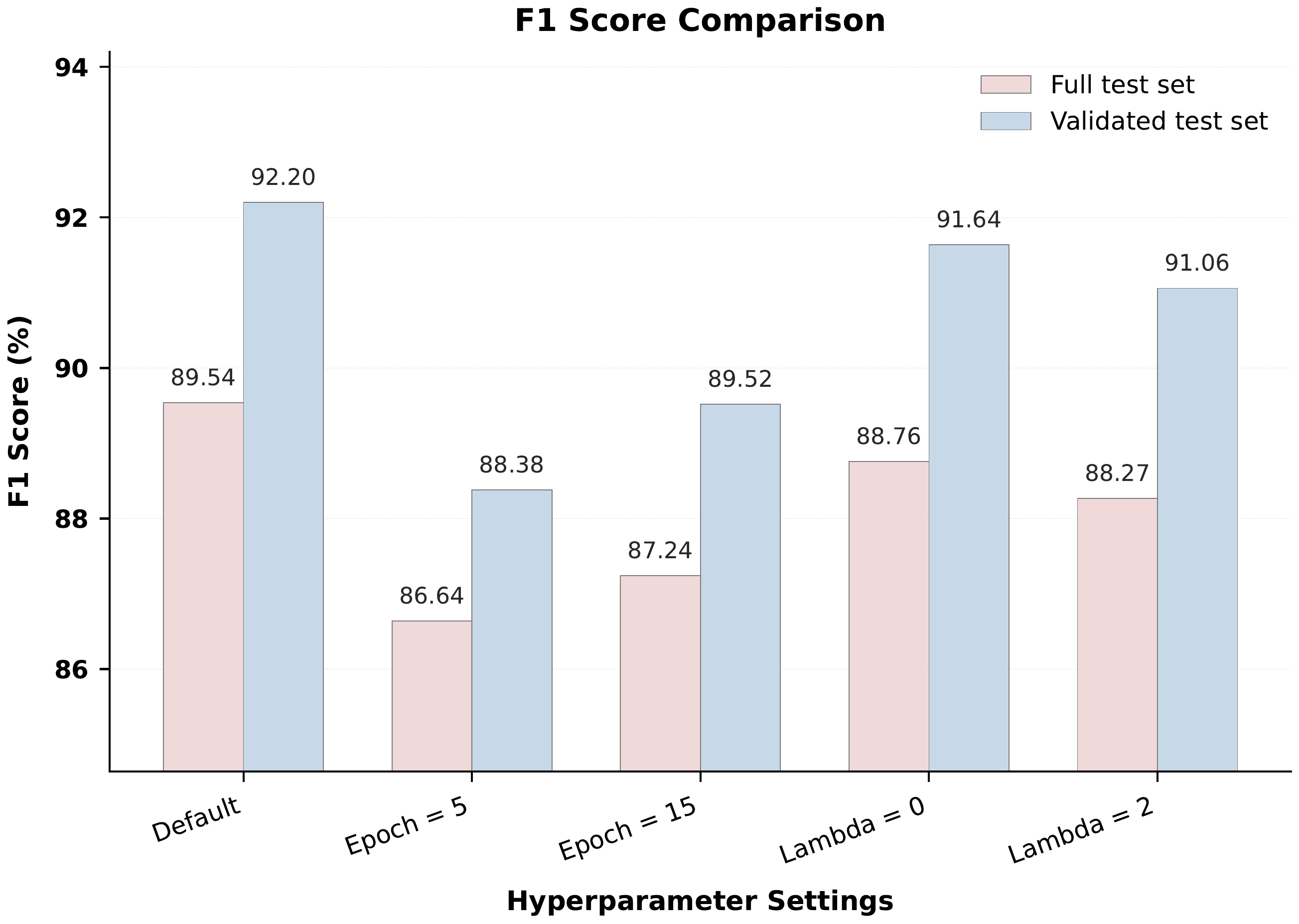}
    \caption{Hyperparameter analysis.}
    \label{fig:ablation1}
    \vspace{-10pt}
\end{figure}

\subsubsection{Motivation}
This RQ investigates the impact of hyperparameter selection in the detector module and the advantage of the iterative enhancement technique on detector performance.
\subsubsection{Approach}
We experiment with various detection model configurations to understand hyperparameter effects on results, specifically adjusting the epoch in fine-tuning and the $\alpha$ in the loss function to balance BCE loss and cosine similarity regularisation. We then examine the impact of different iteration numbers in iterative enhancement.
\subsubsection{Results}
Figure~\ref{fig:ablation1} evaluates how varying the number of epochs, and the lambda parameter affects model accuracy on both the full and validated test sets. With the default settings (epoch = 10, lambda = 1), the model achieves its highest accuracy of 89.54\% on the full test set and 92.2\% on the validated test set. Reducing epochs to 5 lowers the full test set accuracy by 2.90 point and the validated set accuracy by 3.82 point. Increasing epochs to 15 results in an 87.24\% accuracy on the full test set (a 2.30-point drop) and 89.52\% on the validated test set (a 2.68-point reduction).

Adjusting the lambda parameter shows similar effects. Setting lambda to 0 reduces the accuracy on the full test set to 88.76\%, a drop of 0.78 percentage points, and on the validated test set to 91.64\%, a drop of 0.56 percentage points. Increasing lambda to 2 results in an accuracy of 88.27\% on the full test set, a drop of 1.27 percentage points, and 91.06\% on the validated test set, a drop of 1.14 percentage points. These results demonstrate that the default settings for epochs and lambda strike an optimal balance between accuracy and generalization.

Table \ref{tab:ablation2} illustrates the F1 scores of the \detector model across varying iteration numbers, with the default configuration set at 10 iterations. Under this standard condition, the model attains the highest F1 score of 89.54. In comparison, the F1 score is recorded at 86.6 for 0 iterations, indicating a significant enhancement of 2.94 points when transitioning to the default configuration. With 1 iteration, the F1 score exhibits a slight increase to 86.92, reflecting an improvement of 0.32 points over 0 iterations. At the 5-iteration mark, the F1 score attains a value of 87.84, representing a gain of 0.92 points compared to 1 iteration, although still falling short by 1.70 points relative to the default setting. \added{Notably, when the number of iterations is increased to 20, the F1 score registers at 89.42, which is a marginal decrement of 0.12 points compared to the default setting, thereby indicating diminishing returns with the escalation of iterations. This slight performance drop suggests the potential onset of overfitting, underscoring the importance of selecting an optimal number of iterations to balance performance gains with this risk. These findings reinforce the notion that our default configuration offers an optimal setting.}


\begin{table}[tbp]
\centering
\caption{\detector performance with different iteration number.}
\label{tab:ablation2}
\begin{tabular}{@{}cccccc@{}}
\toprule
Iteration number & 0    & 1     & 5     & 10    & 20    \\ \midrule
F1 score         & 86.6 & 86.92 & 87.84 & 89.54 & 89.42 \\ \bottomrule
\end{tabular}
\vspace{-10pt}
\end{table}

\begin{tcolorbox}[boxsep=1pt,left=2pt,right=2pt,top=3pt,bottom=2pt,width=\columnwidth,colback=white!95!black,boxrule=1pt, colbacktitle=white!30!black,toptitle=2pt,bottomtitle=1pt,opacitybacktitle=0.4,fonttitle=\small,fontupper=\small]
\textbf{RQ2 Result:} Experiments show that the model is robust across various hyperparameter configurations. Although default settings offer balanced performance, tuning parameters like epochs and iteration numbers also yields high accuracy and F1 scores, highlighting the model's resilience and adaptability across different scenarios.
\end{tcolorbox}

\vspace{-10pt}
\subsection{RQ3: How effective is \fixer in fixing the inconsistency of the existing code comments?}
\subsubsection{Motivation}
This RQ examines our \fixer's effectiveness in correcting code comment inconsistencies.

\subsubsection{Approach}
To evaluate fix performance, we compare it with existing models, including Panthaplackel et al.~\cite{panthaplackel2021Deepa} and C4RLLaMA~\cite{rong2024code}. For Panthaplackel et al., we use the pre-trained update model due to its superior fix performance. Regarding C4RLLaMA, we adopt the "just-in-time with standard diff" setting, which corresponds with our CCI task in just-in-time mode. We use all inconsistent cases as the test set (i.e. filter out all consistent cases, leaving 1065 cases in the full test set and 150 cases in the validated test set), unlike previous work~\cite{panthaplackel2021Deepa}. This change is important because, in consistent cases, the unchanged old comment matches the ground truth exactly, causing evaluation metrics like BLEU-4 and GLEU to be 1, which inflates the evaluation results.

\subsubsection{Results}
The results presented in Table \ref{tab:fix_result} demonstrate the performance of various fixing models. Notably, \fixer consistently outperforms all other models, delivering exceptional results across all metrics on both the full test set and the validated test set.
In the full test set, \fixer achieves the highest BLEU-4 and METEOR scores of 68.78 and 67.60, respectively, surpassing the next-best model, C4RLLaMA, by margins of 4.22 and 9.34 points. Similarly, in the validated test set, CCIFixer excels further, attaining SARI and GLEU scores of 74.56 and 74.90, which significantly exceed C4RLLaMA’s scores of 67.04 and 66.83, with improvements of 7.52 and 8.07 points, respectively.
For the “No Update” baseline, where the old comment is directly used as the predicted output, relatively high performance is still achieved on certain metrics. This is attributed to the inherent similarity between the inconsistent comments and the ground truth, where minimal adjustments suffice to resolve inconsistencies.

\begin{figure}
    \centering
    \includegraphics[width=0.8\linewidth]{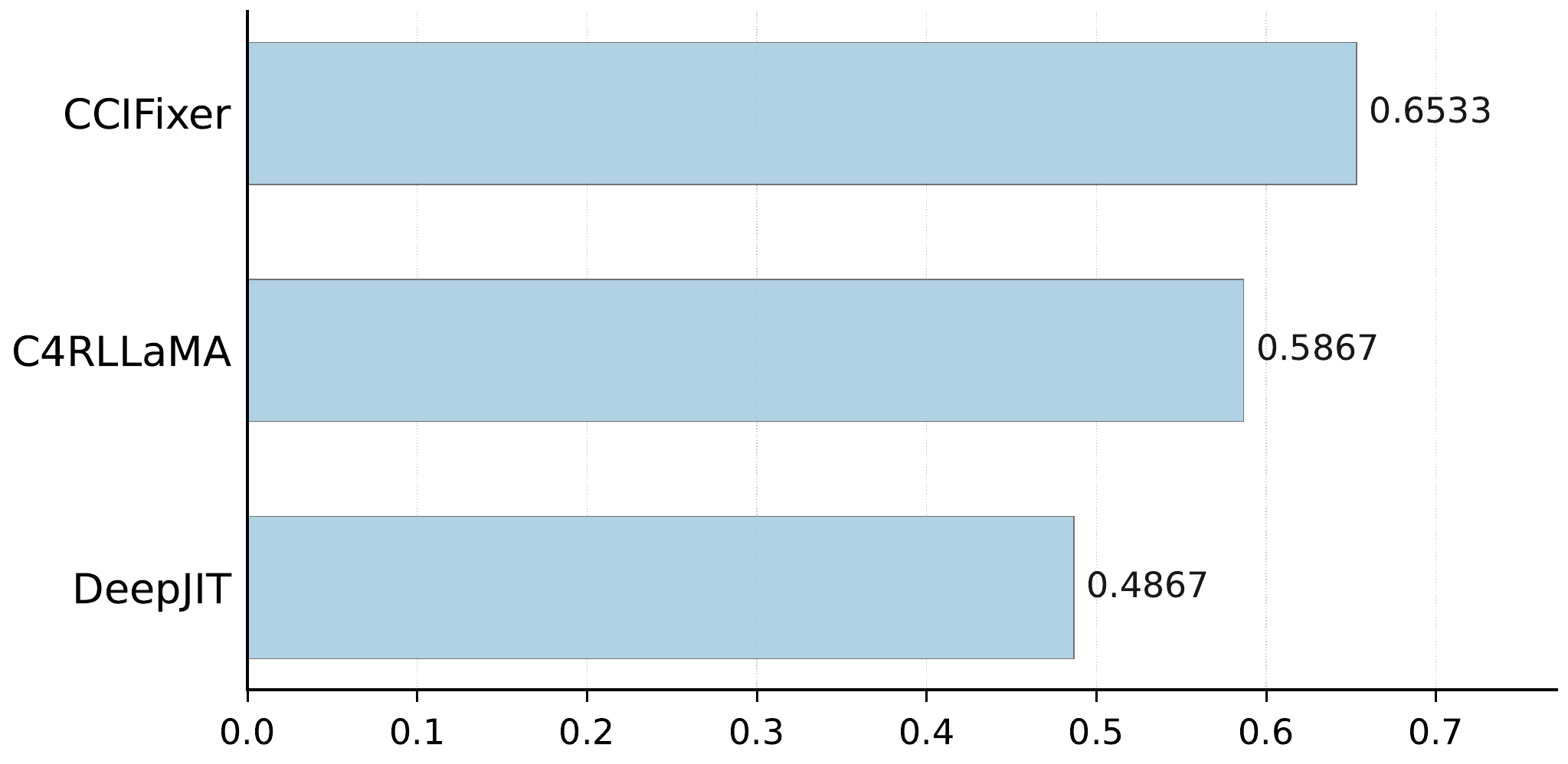}
    \caption{Success rate of human evaluation for the fixing.}
    \label{fig:manual}
    \vspace{-10pt}
\end{figure}

\textit{Human evaluation.} Metrics such as BLEU-4 and METEOR, which emphasize text similarity, provide only a partial evaluation of the fixing efficacy~\cite{rong2024code}. To comprehensively assess our approach, two authors independently examined the predicted results in the validated test set to confirm the accuracy of the fixes. A unanimous consensus between the authors signifies a successful instance. The success rate is calculated as the ratio of successful cases to the total number of fixes. The authors attained a Cohen’s Kappa coefficient of 0.86. Figure~\ref{fig:manual} illustrates the human assessment results for the CCI fixing task. It can be inferred that in comparison to the baselines, \fixer achieves the highest success rate of 0.6533, surpassing the rates of 0.4867 for DeepJIT and 0.5867 for C4RLLaMA.

\begin{tcolorbox}[boxsep=1pt,left=2pt,right=2pt,top=3pt,bottom=2pt,width=\columnwidth,colback=white!95!black,boxrule=1pt, colbacktitle=white!30!black,toptitle=2pt,bottomtitle=1pt,opacitybacktitle=0.4,fonttitle=\small,fontupper=\small]
\textbf{RQ3 Result:} 
\fixer consistently outperforms all other models and increases performance by 18.84\% and 12.07\% on GLEU in full and validated test set, respectively. Furthermore, human evaluation achieves a success rate of 65.33\%, implying the potential for practical use.
\end{tcolorbox}
\begin{table*}[htp]
\centering
\scriptsize
\caption{The result of inconsistent case fixes.}
\label{tab:fix_result}
\begin{tabular}{@{}lcccccccc@{}}
\toprule
 \multirow{2}{*}{Model}     & \multicolumn{4}{c}{Full Test Set} & \multicolumn{4}{c}{Validated Test Set} \\
 & BLEU-4  & METEOR  & SARI  & GLEU  & BLEU-4   & METEOR   & SARI    & GLEU   \\
\midrule
No Update                 & 48.31   & 35.27   & 19.16 & 37.03 & 52.33    & 36.52    & 20.62   & 37.69  \\
Jointly trained SEQ       & 54.25   & 39.98   & 53.55 & 51.33 & 62.11    & 57.77    & 63.51   & 57.89  \\
Jointly trained GRAPH     & 54.74   & 40.26   & 53.68 & 51.81 & 63.33    & 58.74    & 64.02   & 59.02  \\
Jointly trained HYBRID    & 55.20    & 40.87   & 54.14 & 52.47 & 64.08    & 59.21    & 64.34   & 59.47  \\
C4RLLaMA                  & 64.56   & 58.26   & 63.47 & 61.23 & 68.67    & 63.19    & 67.04   & 66.83  \\
CCIFixer                  & \textbf{68.78}   & \textbf{67.60}   & \textbf{65.23} & \textbf{72.77} & \textbf{75.63 }   & \textbf{72.45}    & \textbf{74.56}   & \textbf{74.90}   \\ \bottomrule
\end{tabular}
\vspace{-10pt}
\end{table*}
\subsection{RQ4: How do LLM backbones and tuning process enhance the
performance of \fixer?}
\subsubsection{Motivation}
This RQ investigates the impact of our different design on the fixer model.
\subsubsection{Approach}
We examine this issue from two distinct angles. Initially, we substitute various LLM backbones to determine if our design has consistent effects across different models. Subsequently, we remove the fine-tuning and alignment processes to assess their influence on a particular LLM.

\subsubsection{Results}
\begin{figure}[tbp]
    \centering
    \includegraphics[width=0.95\linewidth]{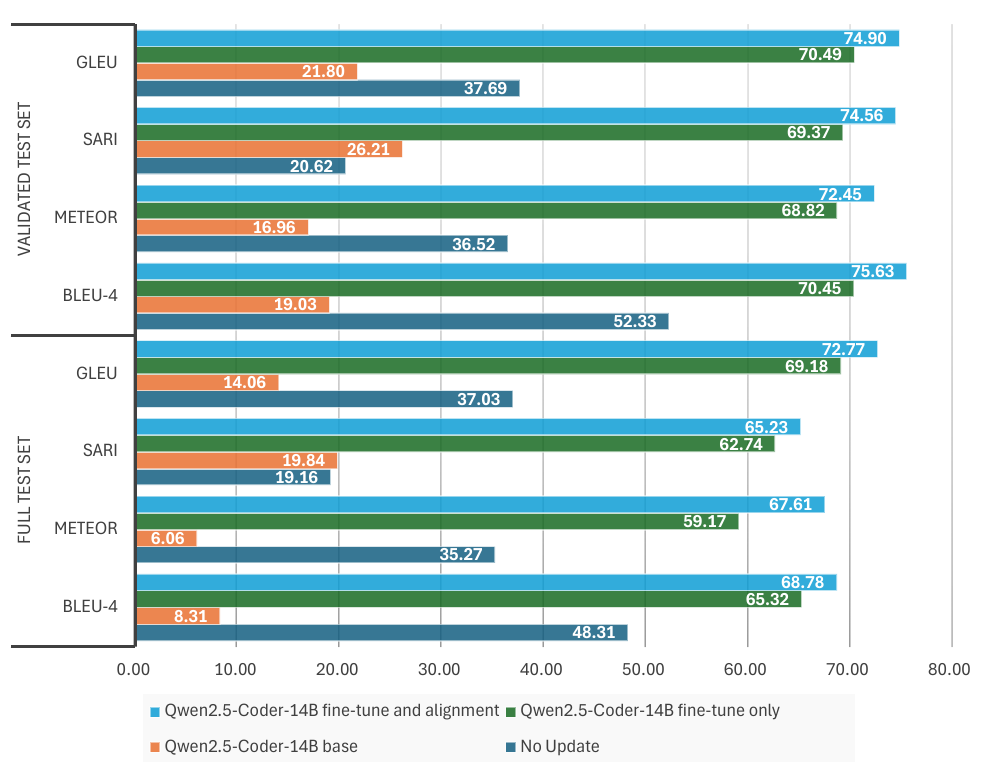}
    \caption{The fixing result of Qwen2.5-Coder-14B with different strategy settings.}
    \label{fig:qwen}
    \vspace{-15pt}
\end{figure}

\begin{figure}[tbp]
    \centering
    \includegraphics[width=0.95\linewidth]{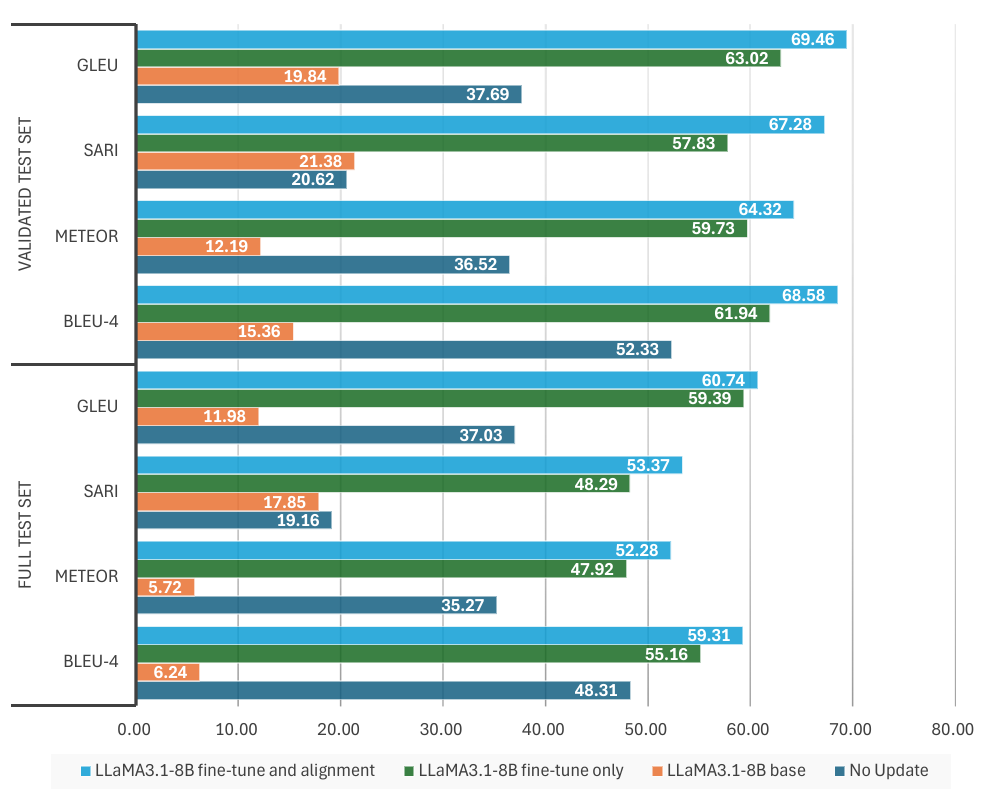}
    \caption{The fixing result of LLaMA3.1-8B with different strategy settings.}
    \label{fig:llama}
    \vspace{-15pt}
\end{figure}

The two figures present the comparative evaluation of Qwen2.5-Coder-14B (Figure~\ref{fig:qwen}) and LLaMA3.1-8B (Figure~\ref{fig:llama}) across multiple configurations (no update, base, fine-tune only, and fine-tune with alignment) using four metrics on both the validated test set and the full test set. First, fine-tuning with alignment achieves the best performance consistently across both models, all metrics, and test sets. Qwen2.5-Coder-14B fine-tuned with alignment achieves GLEU (74.90), while LLaMA3.1-8B achieves GLEU (69.46). This consistent improvement across models validates the effectiveness of the alignment strategy in enhancing model performance. 

Second, models without any fine-tuning perform poorly across all metrics. For example, the base Qwen2.5-Coder-14B achieves GLEU (21.80) and METEOR (16.96) on the validated test set, while LLaMA3.1-8B base scores even lower, with GLEU (19.84) and METEOR (12.19). Interestingly, the result is even worse than ``no update". Note that it does not suggest that a base LLM's fix introduces more inconsistency; it may be because the base LLM lacks the ability to align with the dataset-specific comment preferences. 

Third, Qwen2.5-Coder-14B consistently outperforms LLaMA3.1-8B across all metrics and configurations. For instance, in the validated test set, Qwen2.5-Coder achieves higher GLEU (74.90) and BLEU-4 (75.63) scores compared to LLaMA3.1-8B’s GLEU (69.46) and BLEU-4 (68.58). This performance gap can likely be attributed to the larger parameter size of Qwen2.5-Coder (14B vs. 8B) and its design as a coding-specific model, whereas LLaMA3.1-8B is not primarily trained for code. Nevertheless, the observed improvements across both models demonstrate that the proposed fine-tuning and alignment method is robust and effective, regardless of model scale or type.

\begin{tcolorbox}[boxsep=1pt,left=2pt,right=2pt,top=3pt,bottom=2pt,width=\columnwidth,colback=white!95!black,boxrule=1pt, colbacktitle=white!30!black,toptitle=2pt,bottomtitle=1pt,opacitybacktitle=0.4,fonttitle=\small,fontupper=\small]
\textbf{RQ4 Result:} 
The fine-tuning and alignment techniques enhance the performance of \fixer. Additionally, \fixer is shown to be both robust and efficient for general-purpose LLMs as well as code-focused LLMs.
\end{tcolorbox}

\vspace{-10pt}
\subsection{\added{RQ5: How effective and efficient is \name at end-to-end CCI detection and repair compared to existing method?}}

\subsubsection{\added{Motivation}}

\added{Preious research questions have separately evaluated the performance of our two main component. In real-world application, a developer would first use the detection tool to scan a batch of modified code to identify potential CCIs. Subsequently, they focus their efforts on fixing the cases that are inconsistent.}
\added{Therefore, this RQ aims to evaluate the combined effectiveness and efficiency of \name in a simulated real-world workflow. This directly addresses the need to understand how \detector and \fixer work together and how the system performs when applied to a large-scale codebase, where computational resources and time are practical constraints. We investigate whether our two-stage approach—a lightweight detector followed by a more powerful fixer—is more practical than using a single, monolithic model for both tasks.}

\subsubsection{\added{Approach}}

\added{To evaluate the end-to-end performance in a realistic setting, our approach simulates a complete workflow. We use the entire \bench testset as our test data. We execute the full pipeline of \name, where the \detector component first scans the entire dataset to identify all instances it predicts as inconsistent, after which only these flagged instances are passed to the \fixer component for repair.
We choose C4RLLaMA as a baseline and assess the performance from two deimensions: effectiveness and efficiency. In practice, the time spent on detecting and updating a CCI is significant in the JIT mode. Thus, we introduce the average time spent on detecting and updating a single CCI (\textit{inference time})~\cite{tian2024large} to represent the time efficency.}

\subsubsection{\added{Results}}

\begin{table*}[!t]
\centering
\caption{\added{The end-to-end performance comparisons of \name with the baseline.}}
\label{tab:e2e}
\begin{tabular}{l||cccc|cccc|c}
\toprule
 & Accuracy & Precision & Recall & F1 & BLEU-4 & METEOR & SARI & GLEU &  inference time \\ \midrule
C4RLLaMA & 89.16 & 88.24 & 89.10 & 88.67 & 66.35 &  59.87  & 65.79 & 63.24 & 0.9618 \\
\name & 89.92 & 89.84 & 89.25 & 89.54 & 72.42 & 70.14 & 70.93 & 76.31 & 0.6164   \\ \bottomrule
\end{tabular}
\vspace{-10pt}
\end{table*}

\added{The results of the end-to-end performance comparison between \name and the C4RLLaMA baseline are presented in Table~\ref{tab:e2e}. The findings clearly demonstrate that \name achieves superior performance in both effectiveness and efficiency.}

\added{In terms of effectiveness, \name demonstrates enhancements in both detection and correction tasks. For the detection phase, \detector attains an F1-score of 89.54, surpassing C4RLLaMA's 88.67. The edge is even more substantial in the correction phase, where \fixer significantly outperforms C4RLLaMA in all text generation metrics, achieving a GLEU score of 76.31, while C4RLLaMA is 63.24. This indicates that our proposed pipeline \name not only accurately detects inconsistencies but also delivers superior quality corrections.}

\added{Regarding efficiency, which is critical for application in large codebases with limited resources, \name demonstrates a substantial advantage. The average inference time for \name to process a single instance is 0.6164 seconds, which is approximately 36\% faster than C4RLLaMA's 0.9618 seconds.  This directly validates the architectural design of \name. }

\begin{tcolorbox}[boxsep=1pt,left=2pt,right=2pt,top=3pt,bottom=2pt,width=\columnwidth,colback=white!95!black,boxrule=1pt, colbacktitle=white!30!black,toptitle=2pt,bottomtitle=1pt,opacitybacktitle=0.4,fonttitle=\small,fontupper=\small]
\added{\textbf{RQ5 Result:} 
The end-to-end evaluation confirms that the two-stage architecture of \name is both more effective and time efficient, improving detection and fixing quality while reducing inference time by approximately 36\%, making it a more practical and scalable solution.}
\end{tcolorbox}
\subsection{Case study}
In this section, we conduct a case study to understand the fixing ability of \fixer compared to DeepJIT and C4RLLaMA.

\begin{figure}
    \centering
    \includegraphics[width=0.9\linewidth]{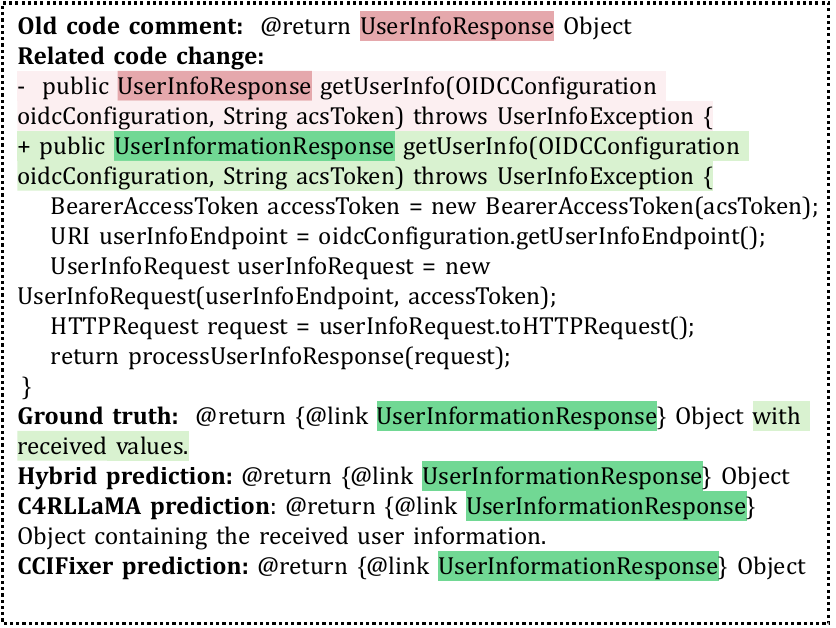}
    \caption{Return inconsistency case.}
    \label{fig:return_case}
    \vspace{-15pt}
\end{figure}

\begin{figure}
    \centering
    \includegraphics[width=0.9\linewidth]{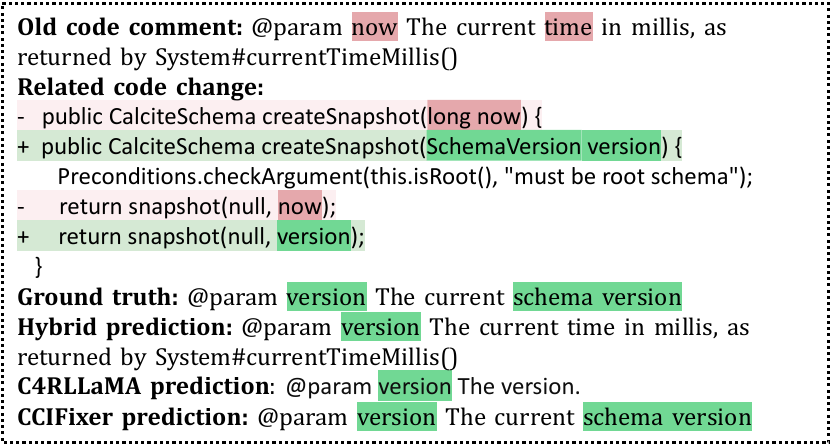}
    \caption{Param inconsistency case.}
    \label{fig:param_case}
    \vspace{-15pt}
\end{figure}

\begin{figure}
    \centering
    \includegraphics[width=0.9\linewidth]{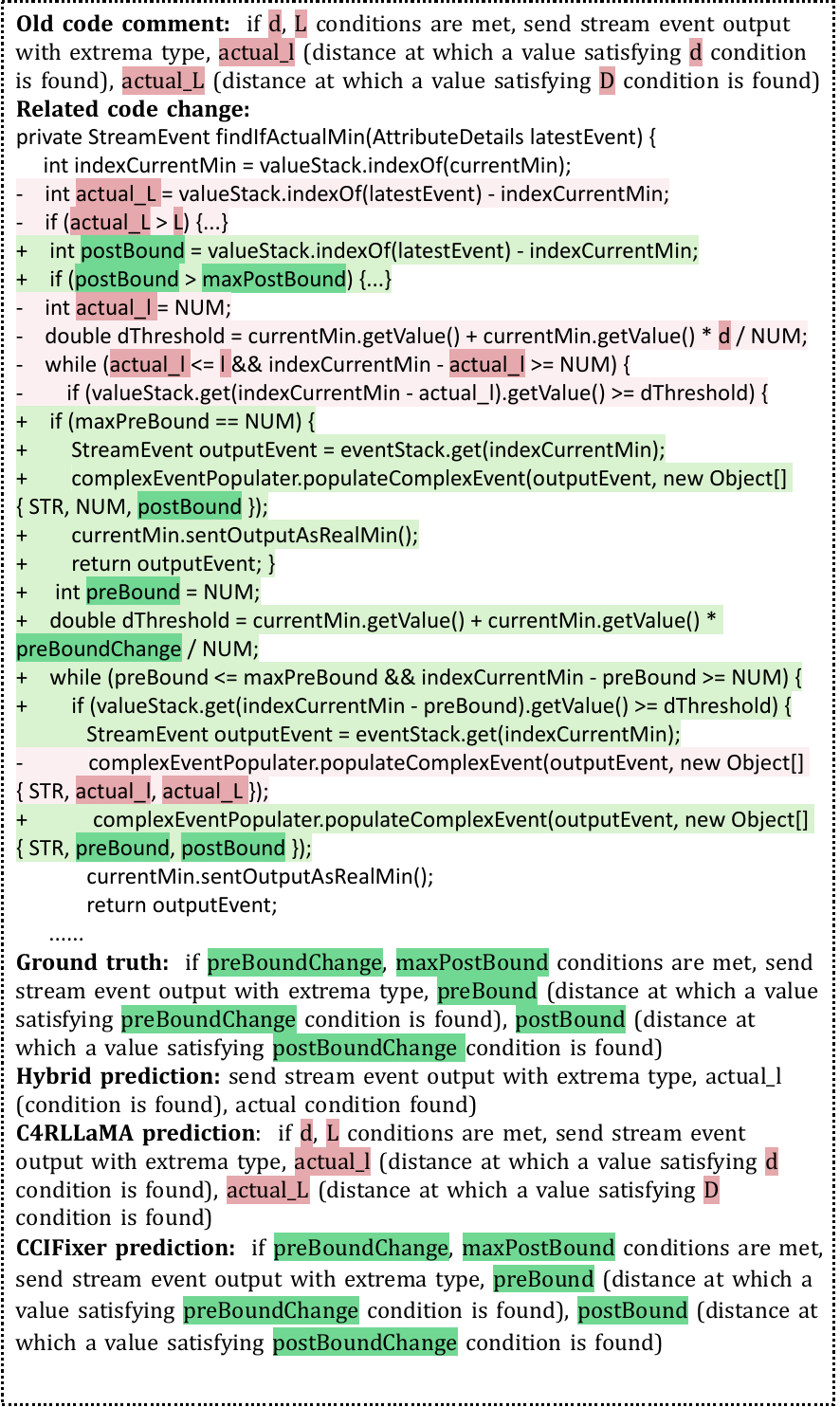}
    \caption{Summary inconsistency case.}
    \label{fig:summary_case}
    \vspace{-15pt}
\end{figure}

Figure~\ref{fig:return_case}--\ref{fig:summary_case} shows three real-world examples from the test sets. \colorbox{lightred}{Light red} and \colorbox{lightgreen}{light green} indicate the code block sections affected by code changes, while \colorbox{darkred}{dark red} and \colorbox{darkgreen}{dark green} highlight the keywords related to code comments. Figure~\ref{fig:return_case} shows a return inconsistent case which changes the return type of the method. Among the three prediction methods, \fixer provides a return comment that is accurate but less informative compared to the ground truth and C4RLLaMA. While it correctly identifies the return type \textit{UserInformationResponse}, it does not offer additional descriptive context about what the object contains. In comparison, C4RLLaMA stands out by including details about the returned object, specifying that it contains “the received user information”, which closely aligns with the ground truth. However, \fixer performs comparably to the Hybrid Prediction, both prioritizing accuracy and brevity but lacking the depth needed for optimal documentation. As a result, \fixer could benefit from incorporating more descriptive elements to enhance the clarity and utility of its predictions.

In the case shown in Figure~\ref{fig:param_case}, \fixer provides the best performance by accurate prediction, which perfectly aligns with the ground truth. It effectively reflects the updated code, capturing the essence of the parameter’s role with clarity and correctness. In contrast, Hybrid Prediction fails to adapt to the code change, erroneously describing the version as “the current time in millis,” which corresponds to the old implementation. This demonstrates a lack of contextual understanding and results in an incorrect prediction. C4RLLaMA’s prediction, ``@param version The version'', is vague and overly generic. While it avoids being outright incorrect, it fails to provide any meaningful detail about the parameter, making it less useful for developers.

The code change updates the logic of the findIfActualMin method by replacing the old actual\_l and actual\_L parameters with preBound and postBound, reflecting a shift in terminology and functionality. Specifically, the condition now uses preBoundChange and maxPostBound instead of d and L, and the output event parameters have been updated accordingly. This indicates a refinement in the method’s logic, aligning it with a more robust schema. Among the fixing methods, \name demonstrates the best performance by accurately reflecting the updated code’s context, correctly identifying preBound and postBound along with their conditions (preBoundChange, maxPostBound) and aligning its comment with the updated logic, matching the ground truth. In contrast, Hybrid Prediction fails to adapt to the code changes, referencing the obsolete actual\_l and providing a vague, incomplete description. C4RLLaMA also fails to adapt, retaining references to outdated parameters and conditions (d, L), making its prediction inaccurate and misleading. \name stands out as the only method providing a clear and contextually appropriate comment.

Overall, the case studies highlight the varying strengths and weaknesses of the three prediction models. \fixer consistently excels in producing accurate and contextually aligned comments that reflect the updated code logic, making it the most reliable model in terms of correctness. However, its comments occasionally lack the depth and descriptiveness needed for optimal clarity, as seen in the return case. Hybrid Prediction shows significant shortcomings, often failing to adapt to code changes and relying on obsolete or incorrect references, which limits its utility. C4RLLaMA, while occasionally more descriptive, suffers from a lack of accuracy and contextual awareness in critical cases, such as retaining outdated parameter names and conditions. These observations underscore \fixer’s strong fixing ability while highlighting the need for further enhancements in providing more descriptive and developer-friendly comments.
        

\begin{figure}[t]
  \centering
  \begin{subfigure}{0.45\textwidth}
    \centering
    \includegraphics[width=\linewidth]{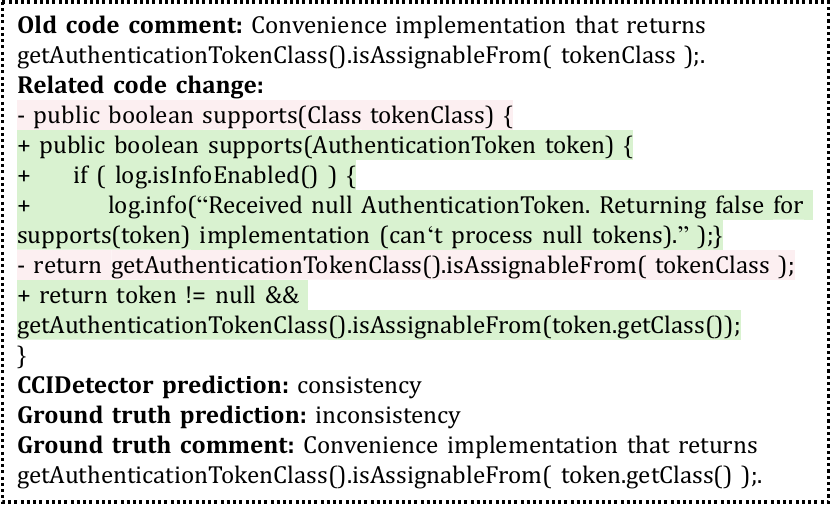}
    \caption{Failure case for \detector.}
    \label{fig:failure_detect}
  \end{subfigure}
  \hfill
  \begin{subfigure}{0.45\textwidth}
    \centering
    \includegraphics[width=\linewidth]{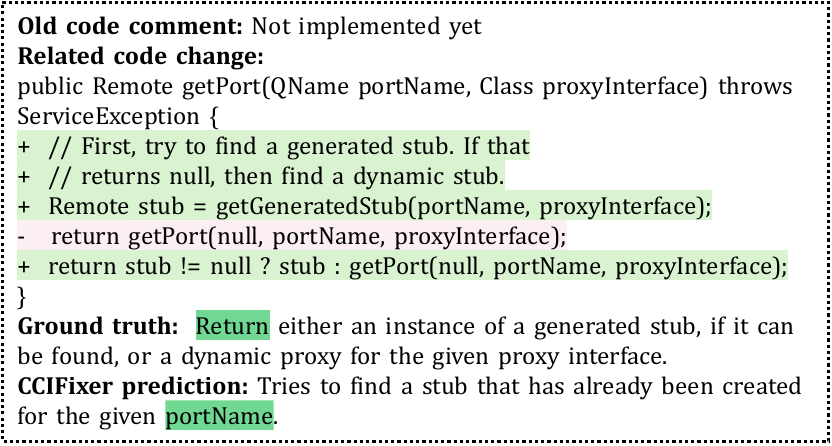}
    \caption{Failure case for \fixer.}
    \label{fig:failure_fix}
  \end{subfigure}
  \caption{\added{Two failure cases in detection and repair stage.}}
  \label{fig:failure}
  \vspace{-15pt}
\end{figure}

\added{While the successful cases demonstrate the effectiveness of \name, analyzing its failures is crucial for a comprehensive understanding of its limitations. Figure~\ref{fig:failure} presents two failure cases, one for \detector and one for \fixer, which reveal areas for future improvement.}

\added{In the \detector failure case shown in Figure~\ref{fig:failure_detect}, the model incorrectly classifies a clear inconsistency as consistent. The code modification involves changing the method parameter's type from \textit{Class} to \textit{AuthenticationToken} and updating the return statement's logic to handle the new type (i.e., \textit{token.getClass()}) along with a null check. \detector's failure to detect this inconsistency likely stems from its inability to fully grasp the semantic shift behind the syntactically similar code. The core API call, isAssignableFrom, remains in both versions of the code. The model may have over-indexed on this syntactic similarity, failing to recognize the critical change in how the parameter is used (\textit{token.getClass()} and \textit{tokenClass}). This suggests that the \detector, while generally robust, can be brittle when faced with changes that are subtle in syntax but significant in semantics, highlighting a need for deeper semantic reasoning capabilities.}

\added{Figure~\ref{fig:failure_fix} illustrates a different kind of limitation for \fixer, one that arises when the original comment lacks substantive content. In this case, the old comment is merely a placeholder, ``\textit{Not implemented yet}", which provides no semantic information about the method’s intended functionality. Consequently, when the method's body undergoes a significant change from being empty to containing a complex implementation with conditional logic, \fixer is deprived of a crucial reference. It cannot perform a comparative analysis between the old comment and new code; instead, it must generate a comment entirely from scratch based solely on its interpretation of the new code.}

\added{
This lack of a content-rich anchor makes the generation task significantly more challenging. Without the context information from a previous comment, the model's output can become less reliable, especially for non-trivial code. As seen in its prediction, ``\textit{Tries to find a stub that has already been created for the given portName}", \fixer only grasps a partial aspect of the new logic. The absence of a meaningful old comment as a guide seems to have hindered its ability to form a complete, holistic understanding of the two-path control flow, leading to an incomplete and inaccurate summary. This failure highlights a specific weakness: \fixer is most effective when it can update an existing description, but its performance diminishes when it must perform pure code summarization on complex logic without any prior semantic context.
}

\vspace{-10pt}
\section{Discussion}
\label{discussion}

\subsection{Implications}
\textit{The necessity of building high-quality datasets.}
\added{To the best of our knowledge, although some previous studies~\cite{xu2023Data,rong2024code} have recognized the possible influence of assumptions made during data collection on labeling errors, this research is pioneering in quantifying the extent of these errors. Our initial analysis of the widely-used dataset reveals that, within our manually inspected sample, 45.67\% of positive instances are incorrectly identified. This contributes considerable noise, potentially degrading the effectiveness of methods trained with these datasets.}


\textit{Appropriate data augmentation can improve performance.} \added{Data augmentation has long been explored in traditional NLP fields~\cite{wei2019eda,shorten2021text}. Inspired by this, recent advancements in LLMs have spurred a new wave of data generation methods. For instance, several studies have successfully employed LLMs to create new training data for various SE tasks~\cite{ubani2023zeroshotdataaug,wang2021want,wei2024magicoder,jainllm,zhong2025LogUpdater}. Our work aligns with this emerging direction. As shown in Table~\ref{tab:ablation2}, our experiments confirm that iteratively synthesizing incorrectly predicted cases could enhance \detector's performance. Nevertheless, owing to constraints on computational resources, we do not evaluate this augmentation technique against other comparative methods. Theoretically, most of the training-dependent approaches could benefit from data augmentation techniques.}

\textit{Text-based metrics cannot indicate the performance of the fix method.}
Table~\ref{tab:fix_result} analyzes the effectiveness of methods in fixing CCI using four text-based metrics. However, we found that higher scores in these metrics do not necessarily reflect better fixing performance. This is partly because pre-fix comments are often already similar to the ground truth, and successful fixes might not improve metric scores if the predicted comments use synonyms. Additionally, different metrics may yield inconsistent results, with models performing well in one but poorly in another. Therefore, metrics like BLEU and METEOR, although widely used in other tasks, might not be ideal for evaluating CCI fixes. An alternative assessment method is a manual evaluation, as demonstrated in RQ4, which addressing these issues, but also has limitations. It may introduce bias and can only be applied to a limited amount of data. Further alternatives still need to be discovered.

\textit{Appropriate design of prompt strategies for LLMs may enhance their performance.}
Our experiments reveal that prompt quality greatly affects the results. Providing LLMs with examples similar to the target code and comment snippets yields better outputs than just using instructions. However, the detection performance with 4-shot ICL remains lower than task-specific baselines. Further research could focus on developing improved prompting strategies.

\added{\textit{Rationale for \detector's design and optimality.} An important consideration is whether the architecture and data enhancement strategy used in \detector are optimal. While the space of all possible models is infinite, our design choices were guided by extensive empirical validation against a wide array of alternatives. In our RQ1 evaluation, \detector is benchmarked against different baselines. Its superior performance across both the full and validated test sets demonstrates the effectiveness of our architectural choices, such as our code diff representation and similarity-based classifier. Furthermore, the contribution of our iterative enhancement strategy is validated in the ablation study for RQ2. The results in Table~\ref{tab:ablation2} show that this targeted approach, which focuses on generating difficult cases the model misclassifies, yields a significant F1-score improvement of 2.94 points compared to no enhancement. Together, these findings confirm that while other configurations may exist, our proposed design is a highly effective and robust solution, validated by its outperformance of numerous alternative strategies.}

\added{
\textit{The efficiency for \name pipeline design.}
\name's architecture is specifically designed for practical efficiency, comprising a lightweight \detector and an LLM-based \fixer. In a real-world workflow, the lightweight detector first filters out the vast majority of consistent code-comment pairs, allowing the computationally expensive fixer to process only the small subset of instances flagged as inconsistent. This two-stage design makes \name approximately 36\% faster than the baseline in the end-to-end evaluation. More importantly, given that the proportion of inconsistencies in real-world code commits is typically far lower than in our test set, \name’s actual efficiency improvement could exceed the measured number, making it a highly practical and scalable solution.
}

\vspace{-5pt}
\subsection{Threats to Validity}
\subsubsection{Construct validity}

\textit{Dataset cleaning.} The process of cleaning datasets is challenging, primarily because fundamental assumptions about existing inconsistencies in the dataset often contain significant flaws. Due to the extensive size of the dataset, manually inspecting numerous inconsistent instances is not feasible. Consequently, we utilized LLMs to autonomously assess inconsistencies, although this approach does not guarantee dataset quality. To mitigate this issue, we introduced a voting mechanism to attest the true inconsistency of retained data. Furthermore, we developed a validated test set to enhance the assessment of performance.
\subsubsection{Internal validity}
\textit{Potential data leakage.} We use LLMs to clean false positive inconsistencies, enhance the dataset by iteratively synthesizing incorrect predictions, and serve as the backbone for correcting inconsistent comments. The training data of LLMs are not publicly available, raising concerns about potential data leakage. However, our experiments show that prompting LLMs to detect inconsistent code comments does not produce promising outcomes. Thus, we believe the fix comments generated by CCIFixer are not merely a result of memorizing training data.
\textit{Human involvement.}
We manually study code-comment inconsistencies, construct a validated test set, and conduct end-to-end performance checks. This process may introduce bias, so we employ two independent annotators with more than seven years of experience to minimize this risk. Any discrepancies are resolved through discussion until consensus is achieved. We achieved a Cohen's Kappa of 0.91 for the manual studies, 0.95 for the test set construction, and 0.86 for the performance checks.

\subsubsection{External validity}

\textit{The selection of programming languages.}
The programming language of our source data is Java, which may raise questions about the generalizability of our proposed method to other programming languages. However, Java is among the most prevalent programming languages for code comment research purposes, in accordance with previous works~\cite{liu2023JustInTime,panthaplackel2021Deepa}. The core idea of detection and fix can be generalized to other languages easily. 
\textit{The selection of LLMs.}
There are no established criteria for evaluating better LLMs. In this study, we choose three popular general-purpose LLMs (GPT3.5, GPT4o, LLaMA3.1) and one code-specific LLM (Deepseek-coder) as baselines, which are commonly used in related software engineering research~\cite{li2024exploring}. Our CCIFixer's performance may vary with different LLM backbones. We choose Qwen2.5-Coder-14b due to its powerful performance in code-related tasks~\cite{qwen2.5_coder_family} and our resource limitations.
\vspace{-10pt}
\section{Related Work}
\label{relatedwork}
\subsection{Code-Comment Inconsistency Detection}
Comments are essential in the software development lifecycle as they facilitate understanding and maintainability of code~\cite{panthaplackel2020Learning}. However, comments are not always consistently updated to reflect modifications in the codebase~\cite{xu2024Code,ratol2017Detectinga}. This lack of synchronization between code and comments can result in inefficiencies, such as wasted time during development and debugging, and may contribute to the introduction of errors in the software~\cite{tan2007icomment}. 

A significant amount of previous research~\cite{hao2023smartcoco,gao2021automating} focused on the issue of inconsistencies in code and comments. These studies can be broadly categorized into two main types based on their detection timing.

\textbf{Just-in-time.} 
Xu et al.~\cite{xu2023Data} tackled data quality concerns, showing that accuracy can improve even with a few label corrections on a small dataset. They proposed an adversarial learning framework that making CCI predictions. Dau et al.~\cite{dau2024DocChecker} introduced DocChecker, a tool that utilizes a pre-trained code language model in conjunction with contrastive learning. Liu et al.~\cite{liu2023JustInTime} developed an obsolete comment detector based on a unique neural network model designed to predict whether a comment requires updating. Steiner et al.~\cite{steiner2022Code} proposed two models—one based on BERT~\cite{devlin2018bert} and the other on Longformer~\cite{beltagy2020longformer}—to detect inconsistencies in the context of natural language inference. Panthaplackel et al.~\cite{panthaplackel2021Deepa} utilized gated graph neural networks to embed code sequences and abstract syntax tree (AST) data, effectively linking comments to code changes.

\textbf{Post-hoc.} 
Rabbi et al.~\cite{rabbi2020ensemble} proposed an ensemble approach using multiple topic modeling techniques to detect inconsistencies between code and comments more robustly.
Ouyang et al.~\cite{ouyang2021prime} explored code-documentation violations in Rust, developing tools to identify and understand breaches in code documentation practices for this specific programming language.
Ratol et al.~\cite{ratol2017Detectinga} utilized heuristic-based rules to detect fragile comments in source code, identifying annotations that are likely to become obsolete due to code changes.

\added{The aforementioned studies represent significant progress in CCI detection and repair, employing techniques from deep learning  to advanced pre-trained models. However, our work is distinguished from these state-of-the-art methods in several aspects. While many existing tools are primarily detectors , \name is a comprehensive framework designed for both detection and automated repair. Our two-stage architecture is fundamentally different from LLM-based solutions. \name combines a lightweight deep learning model (\detector) for efficient, large-scale detection with a powerful LLM-based fixer (\fixer) that is only invoked on the small subset of identified inconsistencies. This design makes our framkework a more practical and scalable solution for real-world software development workflows.}

\vspace{-5pt}
\subsection{Empirical Studies on Code Comment}
Prior research has explored code comments from various perspectives~\cite{linares2015developers,ibrahim2012relationship}. 
For example, Jabrayilzade et al.~\cite{jabrayilzade2024Taxonomy} developed a taxonomy of inline code comment smells and analyzed the frequency of each smell type across software projects. Wang et al.~\cite{wang2023Suboptimal} examined independent comment changes to propose a new proxy for studying suboptimal comments, uncovering insights regarding the prevalence, practices, and impact of commenting guidelines and comment-checking tools. Zhai et al.~\cite{zhai2020CPC} introduced a method leveraging program analysis to systematically derive, refine, and propagate code comments. They developed a comprehensive taxonomy and classifier for comments, demonstrating its effectiveness in enhancing software engineering tasks by generating precise comments and identifying bugs. Wen et al.~\cite{wen2019LargeScale} performed the most extensive empirical study to date on the co-evolution of code and comments, revealing multiple instances where developers either introduced or resolved inconsistencies between code and comments. Their combined quantitative and qualitative analyses provided distilled lessons and actionable recommendations for both researchers and practitioners.
In addition, there is also significant research interest in automated comment generation and improvement~\cite{geng2024Large,lin2022predictive,chen2021my}.

\vspace{-10pt}
\section{Conclusion and Future Work}
\label{conclusion}
\added{To assess data quality issues, our motivating study analyzed a representative sample of the JITDATA dataset, and the results indicated that approximately 45.67\% of the sampled CCIs were false positives.} These mislabeled cases stemmed from four main sources: information additions/deletions, typo corrections, case changes, and lexical modifications. To address these quality concerns, we created \bench, a refined dataset derived from \jitdata. \bench\footnote{Data are available at https://drive.google.com/drive/folders/1c4iTYKoe\\UnXnV9qUXKwp95Fs6eolyLiY?usp=sharing.} incorporates robust de-duplication, syntactic cleaning, and semantic filtering to ensure high-quality data for CCI detection and fixing tasks.
Furthermore, we present \name, an end-to-end framework for handling CCIs. It combines two main components: \detector, a deep learning model for identifying inconsistencies, and \fixer, an LLM-based system for comment repair.

In forthcoming research, our aim will be to rigorously assess methodologies for determining the success of problem resolution. As highlighted previously, metrics derived from text (BLEU) are inadequate for this task. Conversely, manual evaluation is neither scalable for extensive datasets nor free from potential bias. The concept of LLM-as-a-judge has recently emerged as a viable avenue for SE tasks~\cite{wang2025can}. We will persist in investigating the practicality of utilizing this approach for evaluating tasks.

\bibliographystyle{IEEEtran}
\bibliography{sample-base}

\begin{thebibliography}{10}
\providecommand{\url}[1]{#1}
\csname url@samestyle\endcsname
\providecommand{\newblock}{\relax}
\providecommand{\bibinfo}[2]{#2}
\providecommand{\BIBentrySTDinterwordspacing}{\spaceskip=0pt\relax}
\providecommand{\BIBentryALTinterwordstretchfactor}{4}
\providecommand{\BIBentryALTinterwordspacing}{\spaceskip=\fontdimen2\font plus
\BIBentryALTinterwordstretchfactor\fontdimen3\font minus \fontdimen4\font\relax}
\providecommand{\BIBforeignlanguage}[2]{{%
\expandafter\ifx\csname l@#1\endcsname\relax
\typeout{** WARNING: IEEEtran.bst: No hyphenation pattern has been}%
\typeout{** loaded for the language `#1'. Using the pattern for}%
\typeout{** the default language instead.}%
\else
\language=\csname l@#1\endcsname
\fi
#2}}
\providecommand{\BIBdecl}{\relax}
\BIBdecl

\bibitem{pascarella2019classifying}
L.~Pascarella, M.~Bruntink, and A.~Bacchelli, ``Classifying code comments in java software systems,'' \emph{Empirical Software Engineering ({{EMSE}})}, vol.~24, no.~3, pp. 1499--1537, 2019.

\bibitem{padioleau2009listening}
Y.~Padioleau, L.~Tan, and Y.~Zhou, ``Listening to programmers—taxonomies and characteristics of comments in operating system code,'' in \emph{2009 IEEE 31st International Conference on Software Engineering ({{ICSE}})}.\hskip 1em plus 0.5em minus 0.4em\relax IEEE, 2009, pp. 331--341.

\bibitem{buse2009learning}
R.~P. Buse and W.~R. Weimer, ``Learning a metric for code readability,'' \emph{IEEE Transactions on software engineering ({{TSE}})}, vol.~36, no.~4, pp. 546--558, 2009.

\bibitem{zhong2025toward}
R.~Zhong, ``{ Towards Quality Assurance of Natural Language in Code },'' in \emph{2025 IEEE/ACM 47th International Conference on Software Engineering: Companion Proceedings (ICSE-Companion)}.\hskip 1em plus 0.5em minus 0.4em\relax IEEE, May 2025, pp. 187--189.

\bibitem{liu2023JustInTime}
Z.~Liu, X.~Xia, D.~Lo, M.~Yan, and S.~Li, ``Just-{{In-Time Obsolete Comment Detection}} and {{Update}},'' \emph{IEEE Transactions on Software Engineering ({{TSE}})}, vol.~49, no.~1, pp. 1--23, 2023.

\bibitem{springdatamongodb}
S.~Projects, ``Code comment inconsistency cases in spring-data-mongodb,'' \url{https://github.com/spring-projects/spring-data-mongodb/blob/74654cd7c788cdc53b05afe892607396a00f50a2/spring-data-mongodb/src/main/java/org/springframework/data/mongodb/core/mapreduce/MapReduceResults.java#L142}, 2024, accessed: 2024-12-15.

\bibitem{liu2018Automatic}
Z.~Liu, H.~Chen, X.~Chen, X.~Luo, and F.~Zhou, ``Automatic {{Detection}} of {{Outdated Comments During Code Changes}},'' in \emph{2018 {{IEEE}} 42nd {{Annual Computer Software}} and {{Applications Conference}} ({{COMPSAC}})}, vol.~01, 2018, pp. 154--163.

\bibitem{panthaplackel2021Deepa}
S.~Panthaplackel, J.~J. Li, M.~Gligoric, and R.~J. Mooney, ``Deep {{Just-In-Time Inconsistency Detection Between Comments}} and {{Source Code}},'' \emph{Proceedings of the AAAI Conference on Artificial Intelligence ({{AAAI}})}, vol.~35, no.~1, pp. 427--435, 2021.

\bibitem{xu2023Data}
S.~Xu, Y.~Yao, F.~Xu, T.~Gu, J.~Xu, and X.~Ma, ``Data {{Quality Matters}}: {{A Case Study}} of {{Obsolete Comment Detection}},'' in \emph{2023 {{IEEE}}/{{ACM}} 45th {{International Conference}} on {{Software Engineering}} ({{ICSE}})}.\hskip 1em plus 0.5em minus 0.4em\relax Melbourne, Australia: IEEE, 2023, pp. 781--793.

\bibitem{rong2024code}
G.~Rong, Y.~Yu, S.~Liu, X.~Tan, T.~Zhang, H.~Shen, and J.~Hu, ``Code comment inconsistency detection and rectification using a large language model,'' in \emph{2025 IEEE/ACM 47th International Conference on Software Engineering (ICSE)}.\hskip 1em plus 0.5em minus 0.4em\relax IEEE, 2025, pp. 432--443.

\bibitem{xu2024Code}
Z.~Xu, S.~Guo, Y.~Wang, R.~Chen, H.~Li, X.~Li, and H.~Jiang, ``Code {{Comment Inconsistency Detection Based}} on {{Confidence Learning}},'' \emph{IEEE Transactions on Software Engineering ({{TSE}})}, vol.~50, no.~3, pp. 598--617, 2024.

\bibitem{lyu2024automatic}
M.~R. Lyu, B.~Ray, A.~Roychoudhury, S.~H. Tan, and P.~Thongtanunam, ``Automatic programming: Large language models and beyond,'' \emph{ACM Trans. Softw. Eng. Methodol.}, vol.~34, no.~5, May 2025.

\bibitem{OpenAIGPT4o}
\BIBentryALTinterwordspacing
{OpenAI}, ``Gpt-4o,'' 2024, accessed: 2024-06-18. [Online]. Available: \url{https://openai.com/index/hello-gpt-4o/}
\BIBentrySTDinterwordspacing

\bibitem{anthropic2024claude}
\BIBentryALTinterwordspacing
Anthropic, ``Claude 3.5: Sonnet,'' 2024, accessed: 2024-07-10. [Online]. Available: \url{https://www.anthropic.com/news/claude-3-5-sonnet}
\BIBentrySTDinterwordspacing

\bibitem{llama3modelcard}
\BIBentryALTinterwordspacing
AI@Meta, ``Llama 3 model card,'' 2024. [Online]. Available: \url{https://github.com/meta-llama/llama3/blob/main/MODEL_CARD.md}
\BIBentrySTDinterwordspacing

\bibitem{panthaplackel2020associating}
S.~Panthaplackel, M.~Gligoric, R.~J. Mooney, and J.~J. Li, ``Associating natural language comment and source code entities,'' in \emph{Proceedings of the AAAI Conference on Artificial Intelligence ({{AAAI}})}, vol.~34, no.~05, 2020, pp. 8592--8599.

\bibitem{boslaugh2012statistics}
S.~Boslaugh, \emph{Statistics in a nutshell: A desktop quick reference}.\hskip 1em plus 0.5em minus 0.4em\relax O'Reilly Media, Inc., 2012.

\bibitem{sim2005kappa}
J.~Sim and C.~C. Wright, ``The kappa statistic in reliability studies: use, interpretation, and sample size requirements,'' \emph{Physical therapy}, vol.~85, no.~3, pp. 257--268, 2005.

\bibitem{dau2024DocChecker}
A.~T.~V. Dau, J.~L.~C. Guo, and N.~D.~Q. Bui, ``{{DocChecker}}: {{Bootstrapping Code Large Language Model}} for {{Detecting}} and {{Resolving Code-Comment Inconsistencies}},'' \emph{arXiv preprint arXiv:2306.06347}, 2024.

\bibitem{steiner2022Code}
T.~Steiner and R.~Zhang, ``Code comment inconsistency detection with bert and longformer,'' \emph{arXiv preprint arXiv:2207.14444}, 2022.

\bibitem{zheng2023judging}
L.~Zheng, W.-L. Chiang, Y.~Sheng, S.~Zhuang, Z.~Wu, Y.~Zhuang, Z.~Lin, Z.~Li, D.~Li, E.~Xing \emph{et~al.}, ``Judging llm-as-a-judge with mt-bench and chatbot arena,'' \emph{Advances in Neural Information Processing Systems ({{NeurIPS}})}, vol.~36, pp. 46\,595--46\,623, 2023.

\bibitem{livecodebench}
LiveCodeBench, ``Livecodebench leaderboard,'' \url{https://livecodebench.github.io/leaderboard.html}, 2024, accessed: 2024-12-27.

\bibitem{gao2024search}
S.~Gao, C.~Gao, W.~Gu, and M.~Lyu, ``Search-based llms for code optimization,'' in \emph{2025 IEEE/ACM 47th International Conference on Software Engineering (ICSE)}.\hskip 1em plus 0.5em minus 0.4em\relax IEEE, 2025, pp. 254--266.

\bibitem{zhong2025larger}
R.~Zhong, Y.~Li, G.~Yu, W.~Gu, J.~Kuang, Y.~Huo, and M.~R. Lyu, ``Larger is not always better: Exploring small open-source language models in logging statement generation,'' \emph{arXiv preprint arXiv:2505.16590}, 2025.

\bibitem{aroraask}
S.~Arora, A.~Narayan, M.~F. Chen, L.~Orr, N.~Guha, K.~Bhatia, I.~Chami, and C.~Re, ``Ask me anything: A simple strategy for prompting language models,'' in \emph{The Eleventh International Conference on Learning Representations}, 2023.

\bibitem{gao2023makes}
S.~Gao, X.-C. Wen, C.~Gao, W.~Wang, H.~Zhang, and M.~R. Lyu, ``What makes good in-context demonstrations for code intelligence tasks with llms?'' in \emph{2023 38th IEEE/ACM International Conference on Automated Software Engineering (ASE)}.\hskip 1em plus 0.5em minus 0.4em\relax IEEE, 2023, pp. 761--773.

\bibitem{leclair2019neural}
A.~LeClair, S.~Jiang, and C.~McMillan, ``A neural model for generating natural language summaries of program subroutines,'' in \emph{2019 IEEE/ACM 41st International Conference on Software Engineering (ICSE)}.\hskip 1em plus 0.5em minus 0.4em\relax IEEE, 2019, pp. 795--806.

\bibitem{python_difflib}
\BIBentryALTinterwordspacing
P.~S. Foundation, \emph{difflib}, 2025, accessed: 2025-11-27. [Online]. Available: \url{https://docs.python.org/3/library/difflib.html}
\BIBentrySTDinterwordspacing

\bibitem{panthaplackel2020Learning}
S.~Panthaplackel, P.~Nie, M.~Gligoric, J.~J. Li, and R.~Mooney, ``Learning to {{Update Natural Language Comments Based}} on {{Code Changes}},'' in \emph{Proceedings of the 58th {{Annual Meeting}} of the {{Association}} for {{Computational Linguistics}} ({{ACL}})}.\hskip 1em plus 0.5em minus 0.4em\relax Online: Association for Computational Linguistics, 2020, pp. 1853--1868.

\bibitem{guo2022unixcoder}
D.~Guo, S.~Lu, N.~Duan, Y.~Wang, M.~Zhou, and J.~Yin, ``Unixcoder: Unified cross-modal pre-training for code representation,'' in \emph{Proceedings of the 60th Annual Meeting of the Association for Computational Linguistics (Volume 1: Long Papers)}, 2022, pp. 7212--7225.

\bibitem{dong2019unified}
L.~Dong, N.~Yang, W.~Wang, F.~Wei, X.~Liu, Y.~Wang, J.~Gao, M.~Zhou, and H.-W. Hon, ``Unified language model pre-training for natural language understanding and generation,'' \emph{Advances in neural information processing systems ({{NeurIPS}})}, vol.~32, 2019.

\bibitem{zhang2024automatic}
Y.~Zhang, Z.~Qiu, K.-J. Stol, W.~Zhu, J.~Zhu, Y.~Tian, and H.~Liu, ``Automatic commit message generation: A critical review and directions for future work,'' \emph{IEEE Transactions on Software Engineering ({{TSE}})}, 2024.

\bibitem{guo2024exploring}
Q.~Guo, J.~Cao, X.~Xie, S.~Liu, X.~Li, B.~Chen, and X.~Peng, ``Exploring the potential of chatgpt in automated code refinement: An empirical study,'' in \emph{Proceedings of the 46th IEEE/ACM International Conference on Software Engineering (ICSE)}.\hskip 1em plus 0.5em minus 0.4em\relax IEEE, 2024, pp. 1--13.

\bibitem{wei2019eda}
J.~Wei and K.~Zou, ``Eda: Easy data augmentation techniques for boosting performance on text classification tasks,'' in \emph{Proceedings of the 2019 Conference on Empirical Methods in Natural Language Processing and the 9th International Joint Conference on Natural Language Processing (EMNLP-IJCNLP)}, 2019, pp. 6382--6388.

\bibitem{shorten2021text}
C.~Shorten, T.~M. Khoshgoftaar, and B.~Furht, ``Text data augmentation for deep learning,'' \emph{Journal of big Data}, vol.~8, no.~1, p. 101, 2021.

\bibitem{bouzenia2023say}
I.~Bouzenia and M.~Pradel, ``When to say what: Learning to find condition-message inconsistencies,'' in \emph{2023 IEEE/ACM 45th International Conference on Software Engineering (ICSE)}.\hskip 1em plus 0.5em minus 0.4em\relax IEEE, 2023, pp. 868--880.

\bibitem{dai2023auggpt}
H.~Dai, Z.~Liu, W.~Liao, X.~Huang, Y.~Cao, Z.~Wu, L.~Zhao, S.~Xu, F.~Zeng, W.~Liu \emph{et~al.}, ``Auggpt: Leveraging chatgpt for text data augmentation,'' \emph{IEEE Transactions on Big Data}, 2025.

\bibitem{lee2024llm2llm}
N.~Lee, T.~Wattanawong, S.~Kim, K.~Mangalam, S.~Shen, G.~Anumanchipalli, M.~Mahoney, K.~Keutzer, and A.~Gholami, ``Llm2llm: Boosting llms with novel iterative data enhancement,'' in \emph{Findings of the Association for Computational Linguistics ACL 2024}, 2024, pp. 6498--6526.

\bibitem{wang2022self}
Y.~Wang, Y.~Kordi, S.~Mishra, A.~Liu, N.~A. Smith, D.~Khashabi, and H.~Hajishirzi, ``Self-instruct: Aligning language models with self-generated instructions,'' in \emph{Proceedings of the 61st Annual Meeting of the Association for Computational Linguistics (Volume 1: Long Papers)}, 2023, pp. 13\,484--13\,508.

\bibitem{hu2021lora}
E.~J. Hu, P.~Wallis, Z.~Allen-Zhu, Y.~Li, S.~Wang, L.~Wang, W.~Chen \emph{et~al.}, ``Lora: Low-rank adaptation of large language models,'' in \emph{International Conference on Learning Representations}.

\bibitem{ethayarajh2024kto}
K.~Ethayarajh, W.~Xu, N.~Muennighoff, D.~Jurafsky, and D.~Kiela, ``Kto: Model alignment as prospect theoretic optimization,'' \emph{arXiv preprint arXiv:2402.01306}, 2024.

\bibitem{rafailov2024direct}
R.~Rafailov, A.~Sharma, E.~Mitchell, C.~D. Manning, S.~Ermon, and C.~Finn, ``Direct preference optimization: Your language model is secretly a reward model,'' \emph{Advances in Neural Information Processing Systems ({{NeurIPS}})}, vol.~36, 2024.

\bibitem{tian2024large}
Z.~Tian, H.~Shu, D.~Wang, X.~Cao, Y.~Kamei, and J.~Chen, ``Large language models for equivalent mutant detection: How far are we?'' in \emph{Proceedings of the 33rd ACM SIGSOFT International Symposium on Software Testing and Analysis (ISSTA)}, 2024, pp. 1733--1745.

\bibitem{li2024exploring}
Y.~Li, Y.~Huo, Z.~Jiang, R.~Zhong, P.~He, Y.~Su, L.~C. Briand, and M.~R. Lyu, ``Exploring the effectiveness of llms in automated logging statement generation: An empirical study,'' \emph{IEEE Transactions on Software Engineering (TSE)}, vol.~50, no.~12, pp. 3188--3207, 2024.

\bibitem{geng2024Large}
M.~Geng, S.~Wang, D.~Dong, H.~Wang, G.~Li, Z.~Jin, X.~Mao, and X.~Liao, ``Large {{Language Models}} are {{Few-Shot Summarizers}}: {{Multi-Intent Comment Generation}} via {{In-Context Learning}},'' in \emph{Proceedings of the {{IEEE}}/{{ACM}} 46th {{International Conference}} on {{Software Engineering}} ({{ICSE}})}.\hskip 1em plus 0.5em minus 0.4em\relax Lisbon Portugal: IEEE, 2024, pp. 1--13.

\bibitem{feng2020codebert}
Z.~Feng, D.~Guo, D.~Tang, N.~Duan, X.~Feng, M.~Gong, L.~Shou, B.~Qin, T.~Liu, D.~Jiang \emph{et~al.}, ``Codebert: A pre-trained model for programming and natural languages,'' in \emph{Findings of the Association for Computational Linguistics: EMNLP 2020}, 2020, pp. 1536--1547.

\bibitem{gpt-3.5}
\BIBentryALTinterwordspacing
OpenAI., ``Gpt-3.5,'' Mar 2022. [Online]. Available: \url{https://platform.openai.com/docs/models/gpt-3-5}
\BIBentrySTDinterwordspacing

\bibitem{zhu2024deepseek}
Q.~Zhu, D.~Guo, Z.~Shao, D.~Yang, P.~Wang, R.~Xu, Y.~Wu, Y.~Li, H.~Gao, S.~Ma \emph{et~al.}, ``Deepseek-coder-v2: Breaking the barrier of closed-source models in code intelligence,'' \emph{arXiv preprint arXiv:2406.11931}, 2024.

\bibitem{papineni2002bleu}
K.~Papineni, S.~Roukos, T.~Ward, and W.-J. Zhu, ``Bleu: a method for automatic evaluation of machine translation,'' in \emph{Proceedings of the 40th annual meeting of the Association for Computational Linguistics ({{ACL}})}, 2002, pp. 311--318.

\bibitem{banerjee2005meteor}
S.~Banerjee and A.~Lavie, ``Meteor: An automatic metric for mt evaluation with improved correlation with human judgments,'' in \emph{Proceedings of the acl workshop on intrinsic and extrinsic evaluation measures for machine translation and/or summarization}, 2005, pp. 65--72.

\bibitem{xu-etal-2016-optimizing}
W.~Xu, C.~Napoles, E.~Pavlick, Q.~Chen, and C.~Callison-Burch, ``Optimizing statistical machine translation for text simplification,'' \emph{Transactions of the Association for Computational Linguistics}, vol.~4, pp. 401--415, 2016.

\bibitem{napoles2015ground}
C.~Napoles, K.~Sakaguchi, M.~Post, and J.~Tetreault, ``Ground {{Truth}} for {{Grammatical Error Correction Metrics}},'' in \emph{Proceedings of the 53rd {{Annual Meeting}} of the {{Association}} for {{Computational Linguistics}} and the 7th {{International Joint Conference}} on {{Natural Language Processing}} ({{Volume}} 2: {{Short Papers}})}, C.~Zong and M.~Strube, Eds.\hskip 1em plus 0.5em minus 0.4em\relax Beijing, China: Association for Computational Linguistics, 2015, pp. 588--593.

\bibitem{robertson2009probabilistic}
S.~Robertson, H.~Zaragoza \emph{et~al.}, ``The probabilistic relevance framework: Bm25 and beyond,'' \emph{Foundations and Trends{\textregistered} in Information Retrieval}, vol.~3, no.~4, pp. 333--389, 2009.

\bibitem{ubani2023zeroshotdataaug}
S.~Ubani, S.~O. Polat, and R.~Nielsen, ``Zeroshotdataaug: Generating and augmenting training data with chatgpt,'' \emph{arXiv preprint arXiv:2304.14334}, 2023.

\bibitem{wang2021want}
S.~Wang, Y.~Liu, Y.~Xu, C.~Zhu, and M.~Zeng, ``Want to reduce labeling cost? gpt-3 can help,'' in \emph{Findings of the Association for Computational Linguistics: EMNLP 2021}, 2021, pp. 4195--4205.

\bibitem{wei2024magicoder}
Y.~Wei, Z.~Wang, J.~Liu, Y.~Ding, and L.~Zhang, ``Magicoder: empowering code generation with oss-instruct,'' in \emph{Proceedings of the 41st International Conference on Machine Learning (ICML)}, 2024, pp. 52\,632--52\,657.

\bibitem{jainllm}
N.~Jain, T.~Zhang, W.-L. Chiang, J.~E. Gonzalez, K.~Sen, and I.~Stoica, ``Llm-assisted code cleaning for training accurate code generators,'' in \emph{The Twelfth International Conference on Learning Representations (ICLR)}, 2023.

\bibitem{zhong2025LogUpdater}
R.~Zhong, Y.~Li, J.~Kuang, W.~Gu, Y.~Huo, and M.~R. Lyu, ``Logupdater: Automated detection and repair of specific defects in logging statements,'' \emph{ACM Trans. Softw. Eng. Methodol. (TOSEM)}, Apr. 2025.

\bibitem{qwen2.5_coder_family}
\BIBentryALTinterwordspacing
Q.~Team, ``Qwen2.5-coder,'' 2025, accessed: 2025-12-13. [Online]. Available: \url{https://qwenlm.github.io/blog/qwen2.5-coder-family/}
\BIBentrySTDinterwordspacing

\bibitem{ratol2017Detectinga}
I.~K. Ratol and M.~P. Robillard, ``Detecting fragile comments,'' in \emph{2017 32nd {{IEEE}}/{{ACM International Conference}} on {{Automated Software Engineering}} ({{ASE}})}.\hskip 1em plus 0.5em minus 0.4em\relax Urbana, IL: IEEE, 2017, pp. 112--122.

\bibitem{tan2007icomment}
L.~Tan, D.~Yuan, G.~Krishna, and Y.~Zhou, ``/* icomment: Bugs or bad comments?*,'' in \emph{Proceedings of twenty-first ACM SIGOPS symposium on Operating systems principles ({{SOSP}})}, 2007, pp. 145--158.

\bibitem{hao2023smartcoco}
S.~Hao, Y.~Nan, Z.~Zheng, and X.~Liu, ``Smartcoco: Checking comment-code inconsistency in smart contracts via constraint propagation and binding,'' in \emph{2023 38th IEEE/ACM International Conference on Automated Software Engineering (ASE)}.\hskip 1em plus 0.5em minus 0.4em\relax IEEE, 2023, pp. 294--306.

\bibitem{gao2021automating}
Z.~Gao, X.~Xia, D.~Lo, J.~Grundy, and T.~Zimmermann, ``Automating the removal of obsolete todo comments,'' in \emph{Proceedings of the 29th ACM Joint Meeting on European Software Engineering Conference and Symposium on the Foundations of Software Engineering ({{ESEC/FSE}})}, 2021, pp. 218--229.

\bibitem{devlin2018bert}
J.~Devlin, M.-W. Chang, K.~Lee, and K.~Toutanova, ``Bert: Pre-training of deep bidirectional transformers for language understanding,'' in \emph{Proceedings of the 2019 conference of the North American chapter of the association for computational linguistics: human language technologies, volume 1 (long and short papers)}, 2019, pp. 4171--4186.

\bibitem{beltagy2020longformer}
I.~Beltagy, M.~E. Peters, and A.~Cohan, ``Longformer: The long-document transformer,'' \emph{arXiv preprint arXiv:2004.05150}, 2020.

\bibitem{rabbi2020ensemble}
F.~Rabbi, M.~N. Haque, M.~E. Kadir, M.~S. Siddik, and A.~Kabir, ``An ensemble approach to detect code comment inconsistencies using topic modeling.'' in \emph{SEKE}, 2020, pp. 392--395.

\bibitem{ouyang2021prime}
W.~Ouyang and B.~Hua, ``Towards detecting and understanding code-document violations in rust,'' in \emph{2021 IEEE International Symposium on Software Reliability Engineering Workshops (ISSREW)}.\hskip 1em plus 0.5em minus 0.4em\relax IEEE, 2021, pp. 189--197.

\bibitem{linares2015developers}
M.~Linares-V{\'a}squez, B.~Li, C.~Vendome, and D.~Poshyvanyk, ``How do developers document database usages in source code?(n),'' in \emph{2015 30th IEEE/ACM International Conference on Automated Software Engineering (ASE)}.\hskip 1em plus 0.5em minus 0.4em\relax IEEE, 2015, pp. 36--41.

\bibitem{ibrahim2012relationship}
W.~M. Ibrahim, N.~Bettenburg, B.~Adams, and A.~E. Hassan, ``On the relationship between comment update practices and software bugs,'' \emph{Journal of Systems and Software ({{JSS}})}, vol.~85, no.~10, pp. 2293--2304, 2012.

\bibitem{jabrayilzade2024Taxonomy}
E.~Jabrayilzade, A.~Yurto{\u g}lu, and E.~T{\"u}z{\"u}n, ``Taxonomy of inline code comment smells,'' \emph{Empirical Software Engineering ({{EMSE}})}, vol.~29, no.~3, p.~58, 2024.

\bibitem{wang2023Suboptimal}
C.~Wang, H.~He, U.~Pal, D.~Marinov, and M.~Zhou, ``Suboptimal {{Comments}} in {{Java Projects}}: {{From Independent Comment Changes}} to {{Commenting Practices}},'' \emph{ACM Transactions on Software Engineering and Methodology ({{TOSEM}})}, vol.~32, no.~2, pp. 1--33, 2023.

\bibitem{zhai2020CPC}
J.~Zhai, X.~Xu, Y.~Shi, G.~Tao, M.~Pan, S.~Ma, L.~Xu, W.~Zhang, L.~Tan, and X.~Zhang, ``{{CPC}}: Automatically classifying and propagating natural language comments via program analysis,'' in \emph{Proceedings of the {{ACM}}/{{IEEE}} 42nd {{International Conference}} on {{Software Engineering}} ({{ICSE}})}.\hskip 1em plus 0.5em minus 0.4em\relax Seoul South Korea: IEEE, 2020, pp. 1359--1371.

\bibitem{wen2019LargeScale}
F.~Wen, C.~Nagy, G.~Bavota, and M.~Lanza, ``A {{Large-Scale Empirical Study}} on {{Code-Comment Inconsistencies}},'' in \emph{2019 {{IEEE}}/{{ACM}} 27th {{International Conference}} on {{Program Comprehension}} ({{ICPC}})}.\hskip 1em plus 0.5em minus 0.4em\relax Montreal, QC, Canada: IEEE, 2019, pp. 53--64.

\bibitem{lin2022predictive}
B.~Lin, S.~Wang, Z.~Liu, X.~Xia, and X.~Mao, ``Predictive comment updating with heuristics and ast-path-based neural learning: A two-phase approach,'' \emph{IEEE Transactions on Software Engineering ({{TSE}})}, vol.~49, no.~4, pp. 1640--1660, 2022.

\bibitem{chen2021my}
Q.~Chen, X.~Xia, H.~Hu, D.~Lo, and S.~Li, ``Why my code summarization model does not work: Code comment improvement with category prediction,'' \emph{ACM Transactions on Software Engineering and Methodology (TOSEM)}, vol.~30, no.~2, pp. 1--29, 2021.

\bibitem{wang2025can}
R.~Wang, J.~Guo, C.~Gao, G.~Fan, C.~Y. Chong, and X.~Xia, ``Can llms replace human evaluators? an empirical study of llm-as-a-judge in software engineering,'' \emph{arXiv preprint arXiv:2502.06193}, 2025.

\end{thebibliography}

\newpage

\appendices
\section{}

\begin{figure}[h]
    \centering
    \includegraphics[width=0.7\linewidth]{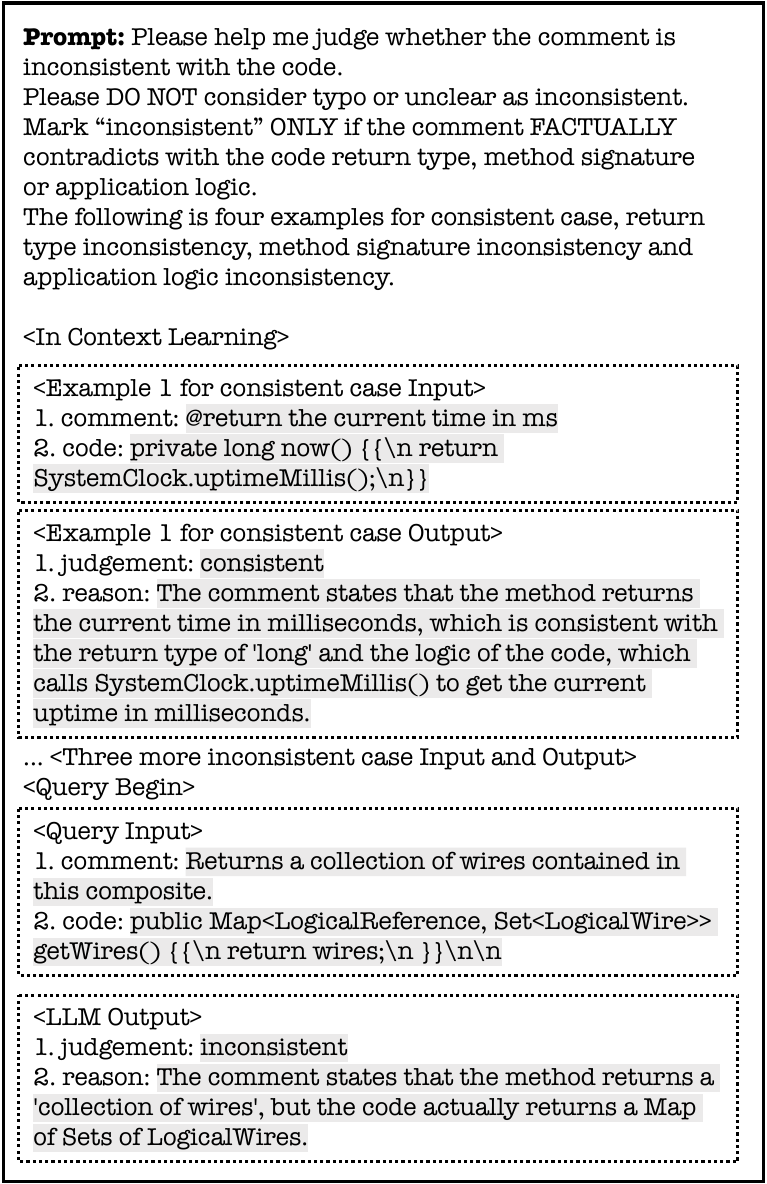}
    \caption{The prompt template for semantic-based filtering.}
    \label{fig:data_prompt}
\end{figure}

\begin{figure}[h]
    \centering
    \includegraphics[width=0.7\linewidth]{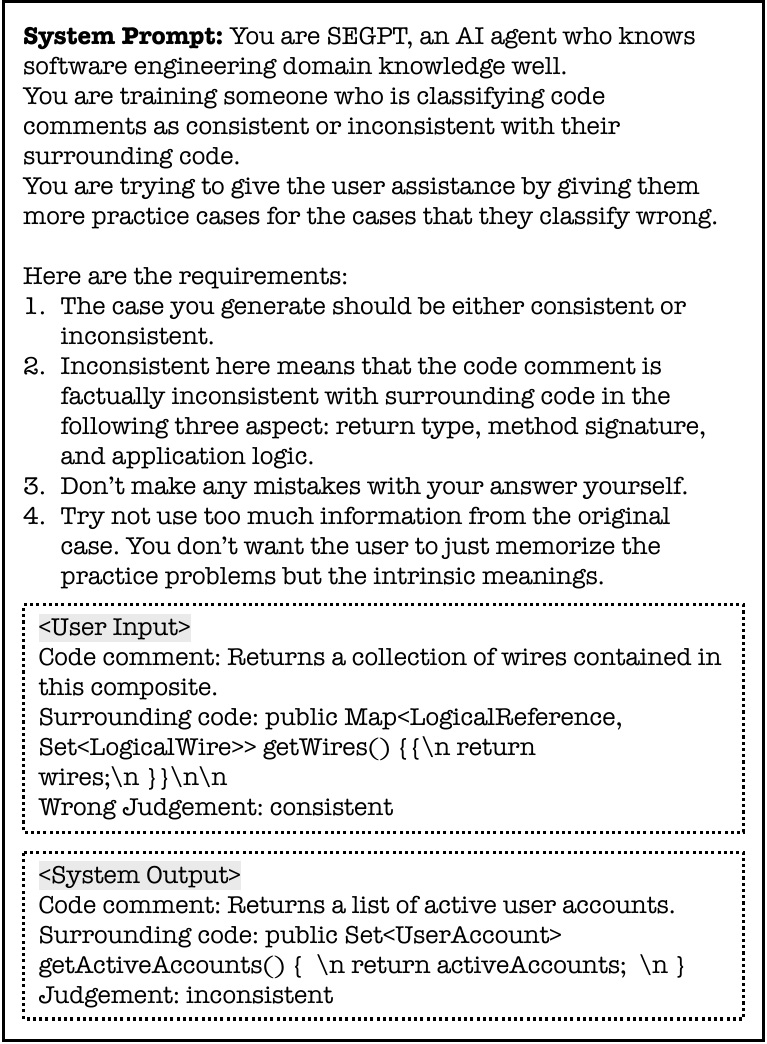}
    \caption{The prompt template for iterative enhancement.}
    \label{fig:enhance_prompt}
\end{figure}

\end{document}